
%
%
\def\gronk#1#2#3
    {\midinsert
         \noindent\hfil\hbox to #1truein
              {\vbox to #2truein
                   {\special{include(#3 origin nocarriage)}\vfill}
              \hfil}
    \hfil\endinsert}
%
%
\expandafter\ifx\csname phyzzx\endcsname\relax\else
 \errhelp{Hit <CR> and go ahead.}
 \errmessage{PHYZZX macros are already loaded or input. }
 \endinput \fi
\catcode`\@=11 
%
%
%
\font\seventeenrm=cmr10 scaled\magstep3
\font\fourteenrm=cmr10 scaled\magstep2
\font\twelverm=cmr10 scaled\magstep1
\font\ninerm=cmr9            \font\sixrm=cmr6

\font\fourteenbf=cmbx10 scaled\magstep2
\font\twelvebf=cmbx10 scaled\magstep1
\font\ninebf=cmbx9            \font\sixbf=cmbx6
\font\seventeeni=cmmi10 scaled\magstep3     \skewchar\seventeeni='177
\font\fourteeni=cmmi10 scaled\magstep2      \skewchar\fourteeni='177
\font\twelvei=cmmi10 scaled\magstep1        \skewchar\twelvei='177
\font\ninei=cmmi9                           \skewchar\ninei='177
\font\sixi=cmmi6                            \skewchar\sixi='177
\font\seventeensy=cmsy10 scaled\magstep3    \skewchar\seventeensy='60
\font\fourteensy=cmsy10 scaled\magstep2     \skewchar\fourteensy='60
\font\twelvesy=cmsy10 scaled\magstep1       \skewchar\twelvesy='60
\font\ninesy=cmsy9                          \skewchar\ninesy='60
\font\sixsy=cmsy6                           \skewchar\sixsy='60

\font\fourteenex=cmex10 scaled\magstep2
\font\twelveex=cmex10 scaled\magstep1
%

\font\fourteensl=cmsl10 scaled\magstep2
\font\twelvesl=cmsl10 scaled\magstep1
\font\ninesl=cmsl9

\font\fourteenit=cmti10 scaled\magstep2
\font\twelveit=cmti10 scaled\magstep1
\font\nineit=cmti9
\font\fourteentt=cmtt10 scaled\magstep2
\font\twelvett=cmtt10 scaled\magstep1
\font\fourteencp=cmcsc10 scaled\magstep2
\font\twelvecp=cmcsc10 scaled\magstep1
\font\tencp=cmcsc10
\newfam\cpfam
\newdimen\b@gheight        \b@gheight=12pt
\newcount\f@ntkey        \f@ntkey=0
\def\f@m{\afterassignment\samef@nt\f@ntkey=}
\def\samef@nt{\fam=\f@ntkey \the\textfont\f@ntkey\relax}
\def\rm{\f@m0 }
\def\mit{\f@m1 }         
\def\cal{\f@m2 }
\def\it{\f@m\itfam}
\def\sl{\f@m\slfam}
\def\bf{\f@m\bffam}
\def\tt{\f@m\ttfam}
\def\caps{\f@m\cpfam}
\def\fourteenpoint{\relax
    \textfont0=\fourteenrm          \scriptfont0=\tenrm
      \scriptscriptfont0=\sevenrm
    \textfont1=\fourteeni           \scriptfont1=\teni
      \scriptscriptfont1=\seveni
    \textfont2=\fourteensy          \scriptfont2=\tensy
      \scriptscriptfont2=\sevensy
    \textfont3=\fourteenex          \scriptfont3=\twelveex
      \scriptscriptfont3=\tenex
    \textfont\itfam=\fourteenit     \scriptfont\itfam=\tenit
    \textfont\slfam=\fourteensl     \scriptfont\slfam=\tensl
    \textfont\bffam=\fourteenbf     \scriptfont\bffam=\tenbf
      \scriptscriptfont\bffam=\sevenbf
    \textfont\ttfam=\fourteentt
    \textfont\cpfam=\fourteencp
    \samef@nt
    \b@gheight=14pt
    \setbox\strutbox=\hbox{\vrule height 0.85\b@gheight
                depth 0.35\b@gheight width\z@ }}
\def\twelvepoint{\relax
    \textfont0=\twelverm          \scriptfont0=\ninerm
      \scriptscriptfont0=\sixrm
    \textfont1=\twelvei           \scriptfont1=\ninei
      \scriptscriptfont1=\sixi
    \textfont2=\twelvesy           \scriptfont2=\ninesy
      \scriptscriptfont2=\sixsy
    \textfont3=\twelveex          \scriptfont3=\tenex
      \scriptscriptfont3=\tenex
    \textfont\itfam=\twelveit     \scriptfont\itfam=\nineit
    \textfont\slfam=\twelvesl     \scriptfont\slfam=\ninesl
    \textfont\bffam=\twelvebf     \scriptfont\bffam=\ninebf
      \scriptscriptfont\bffam=\sixbf
    \textfont\ttfam=\twelvett
    \textfont\cpfam=\twelvecp
    \samef@nt
    \b@gheight=12pt
    \setbox\strutbox=\hbox{\vrule height 0.85\b@gheight
                depth 0.35\b@gheight width\z@ }}
\def\tenpoint{\relax
    \textfont0=\tenrm          \scriptfont0=\sevenrm
      \scriptscriptfont0=\fiverm
    \textfont1=\teni           \scriptfont1=\seveni
      \scriptscriptfont1=\fivei
    \textfont2=\tensy          \scriptfont2=\sevensy
      \scriptscriptfont2=\fivesy
    \textfont3=\tenex          \scriptfont3=\tenex
      \scriptscriptfont3=\tenex
    \textfont\itfam=\tenit     \scriptfont\itfam=\seveni
    \textfont\slfam=\tensl     \scriptfont\slfam=\sevenrm
    \textfont\bffam=\tenbf     \scriptfont\bffam=\sevenbf
      \scriptscriptfont\bffam=\fivebf
    \textfont\ttfam=\tentt
    \textfont\cpfam=\tencp
    \samef@nt
    \b@gheight=10pt
    \setbox\strutbox=\hbox{\vrule height 0.85\b@gheight
                depth 0.35\b@gheight width\z@ }}
%
%
%
\normalbaselineskip = 20pt plus 0.2pt minus 0.1pt
\normallineskip = 1.5pt plus 0.1pt minus 0.1pt
\normallineskiplimit = 1.5pt
\newskip\normaldisplayskip
\normaldisplayskip = 20pt plus 5pt minus 10pt
\newskip\normaldispshortskip
\normaldispshortskip = 6pt plus 5pt
\newskip\normalparskip
\normalparskip = 6pt plus 2pt minus 1pt
\newskip\skipregister
\skipregister = 5pt plus 2pt minus 1.5pt
\newif\ifsingl@    \newif\ifdoubl@
\newif\iftwelv@    \twelv@true
\def\singlespace{\singl@true\doubl@false\spaces@t}
\def\doublespace{\singl@false\doubl@true\spaces@t}
\def\normalspace{\singl@false\doubl@false\spaces@t}
\def\Tenpoint{\tenpoint\twelv@false\spaces@t}
\def\Twelvepoint{\twelvepoint\twelv@true\spaces@t}
\def\spaces@t{\relax
      \iftwelv@ \ifsingl@\subspaces@t3:4;\else\subspaces@t1:1;\fi
       \else \ifsingl@\subspaces@t3:5;\else\subspaces@t4:5;\fi \fi
      \ifdoubl@ \multiply\baselineskip by 5
         \divide\baselineskip by 4 \fi }
\def\subspaces@t#1:#2;{
      \baselineskip = \normalbaselineskip
      \multiply\baselineskip by #1 \divide\baselineskip by #2
      \lineskip = \normallineskip
      \multiply\lineskip by #1 \divide\lineskip by #2
      \lineskiplimit = \normallineskiplimit
      \multiply\lineskiplimit by #1 \divide\lineskiplimit by #2
      \parskip = \normalparskip
      \multiply\parskip by #1 \divide\parskip by #2
      \abovedisplayskip = \normaldisplayskip
      \multiply\abovedisplayskip by #1 \divide\abovedisplayskip by #2
      \belowdisplayskip = \abovedisplayskip
      \abovedisplayshortskip = \normaldispshortskip
      \multiply\abovedisplayshortskip by #1
        \divide\abovedisplayshortskip by #2
      \belowdisplayshortskip = \abovedisplayshortskip
      \advance\belowdisplayshortskip by \belowdisplayskip
      \divide\belowdisplayshortskip by 2
      \smallskipamount = \skipregister
      \multiply\smallskipamount by #1 \divide\smallskipamount by #2
      \medskipamount = \smallskipamount \multiply\medskipamount by 2
      \bigskipamount = \smallskipamount \multiply\bigskipamount by 4 }
\def\normalbaselines{ \baselineskip=\normalbaselineskip
   \lineskip=\normallineskip \lineskiplimit=\normallineskip
   \iftwelv@\else \multiply\baselineskip by 4 \divide\baselineskip by 5
     \multiply\lineskiplimit by 4 \divide\lineskiplimit by 5
     \multiply\lineskip by 4 \divide\lineskip by 5 \fi }
\Twelvepoint  
\interlinepenalty=50
\interfootnotelinepenalty=5000
\predisplaypenalty=9000
\postdisplaypenalty=500
\hfuzz=1pt
\vfuzz=0.2pt
\voffset=0pt
\dimen\footins=8 truein
%
%
%
\def\pagecontents{
   \ifvoid\topins\else\unvbox\topins\vskip\skip\topins\fi
   \dimen@ = \dp255 \unvbox255
   \ifvoid\footins\else\vskip\skip\footins\footrule\unvbox\footins\fi
   \ifr@ggedbottom \kern-\dimen@ \vfil \fi }
\def\makeheadline{\vbox to 0pt{ \skip@=\topskip
      \advance\skip@ by -12pt \advance\skip@ by -2\normalbaselineskip
      \vskip\skip@ \line{\vbox to 12pt{}\the\headline} \vss
      }\nointerlineskip}
\def\makefootline{\baselineskip = 1.5\normalbaselineskip
                 \line{\the\footline}}
\newif\iffrontpage
\newif\ifletterstyle
\newif\ifp@genum
\def\nopagenumbers{\p@genumfalse}
\def\pagenumbers{\p@genumtrue}
\pagenumbers
\newtoks\paperheadline
\newtoks\letterheadline
\newtoks\paperfootline
\newtoks\letterfootline
\newtoks\letterinfo
\newtoks\Letterinfo
\newtoks\date
\footline={\ifletterstyle\the\letterfootline\else\the\paperfootline\fi}
\paperfootline={\hss\iffrontpage\else\ifp@genum\tenrm\folio\hss\fi\fi}
\letterfootline={\iffrontpage\the\letterinfo\else\hfil\fi}
\Letterinfo={\hfil}
\letterinfo={\hfil}
\headline={\ifletterstyle\the\letterheadline\else\the\paperheadline\fi}
\paperheadline={\hfil}
\letterheadline{\iffrontpage \hfil \else
    \rm \ifp@genum page \ \folio\fi \hfil\the\date \fi}
\def\monthname{\relax\ifcase\month 0/\or January\or February\or
   March\or April\or May\or June\or July\or August\or September\or
   October\or November\or December\else\number\month/\fi}
\def\today{\monthname\ \number\day, \number\year}
\date={\today}
\countdef\pageno=1      \countdef\pagen@=0
\countdef\pagenumber=1  \pagenumber=1
\def\advancepageno{\global\advance\pagen@ by 1
   \ifnum\pagenumber<0 \global\advance\pagenumber by -1
    \else\global\advance\pagenumber by 1 \fi \global\frontpagefalse }
\def\folio{\ifnum\pagenumber<0 \romannumeral-\pagenumber
           \else \number\pagenumber \fi }
\def\footrule{\dimen@=\prevdepth\nointerlineskip
   \vbox to 0pt{\vskip -0.25\baselineskip \hrule width 0.35\hsize \vss}
   \prevdepth=\dimen@ }
\newtoks\foottokens
\foottokens={}
\newdimen\footindent
\footindent=24pt
\def\vfootnote#1{\insert\footins\bgroup
   \interlinepenalty=\interfootnotelinepenalty \floatingpenalty=20000
   \singl@true\doubl@false\Tenpoint
   \splittopskip=\ht\strutbox \boxmaxdepth=\dp\strutbox
   \leftskip=\footindent \rightskip=\z@skip
   \parindent=0.5\footindent \parfillskip=0pt plus 1fil
   \spaceskip=\z@skip \xspaceskip=\z@skip
   \the\foottokens
   \Textindent{$ #1 $}\footstrut\futurelet\next\fo@t}
\def\Textindent#1{\noindent\llap{#1\enspace}\ignorespaces}
\def\footnote#1{\attach{#1}\vfootnote{#1}}

\let\footsymbol=\star
\newcount\lastf@@t           \lastf@@t=-1
\newcount\footsymbolcount    \footsymbolcount=0
\newif\ifPhysRev
\def\bumpfootsymbolcount{\relax
   \iffrontpage \bumpfootsymbolNP \else \advance\lastf@@t by 1
     \ifPhysRev \bumpfootsymbolPR \else \bumpfootsymbolNP \fi \fi
   \global\lastf@@t=\pagen@ }
\def\bumpfootsymbolNP{\ifnum\footsymbolcount <0 \global\footsymbolcount =0 \fi
    \ifnum\lastf@@t<\pagen@ \global\footsymbolcount=0
     \else \global\advance\footsymbolcount by 1 \fi }
\def\bumpfootsymbolPR{\ifnum\footsymbolcount >0 \global\footsymbolcount =0 \fi
      \global\advance\footsymbolcount by -1 }
\def\fd@f#1 {\xdef\footsymbol{\mathchar"#1 }}
\def\generatefootsymbol{\ifcase\footsymbolcount \fd@f 13F \or \fd@f 279
    \or \fd@f 27A \or \fd@f 278 \or \fd@f 27B \else
    \ifnum\footsymbolcount <0 \fd@f{023 \number-\footsymbolcount }
     \else \fd@f 203 {\loop \ifnum\footsymbolcount >5
        \fd@f{203 \footsymbol } \advance\footsymbolcount by -1
        \repeat }\fi \fi }

\def\nonfrenchspacing{\sfcode`\.=3001 \sfcode`\!=3000 \sfcode`\?=3000
    \sfcode`\:=2000 \sfcode`\;=1500 \sfcode`\,=1251 }
\nonfrenchspacing
\newdimen\d@twidth
{\setbox0=\hbox{s.} \global\d@twidth=\wd0 \setbox0=\hbox{s}
    \global\advance\d@twidth by -\wd0 }
\def\removehglue{\loop \unskip \ifdim\lastskip >\z@ \repeat }
\def\roll@ver#1{\removehglue \nobreak \count255 =\spacefactor \dimen@=\z@
    \ifnum\count255 =3001 \dimen@=\d@twidth \fi
    \ifnum\count255 =1251 \dimen@=\d@twidth \fi
    \iftwelv@ \kern-\dimen@ \else \kern-0.83\dimen@ \fi
   #1\spacefactor=\count255 }
\def\step@ver#1{\relax \ifmmode #1\else \ifhmode
    \roll@ver{${}#1$}\else {\setbox0=\hbox{${}#1$}}\fi\fi }
\def\attach#1{\step@ver{\strut^{\mkern 2mu #1} }}
%
%
%
\newcount\chapternumber      \chapternumber=0
\newcount\sectionnumber      \sectionnumber=0
\newcount\equanumber         \equanumber=0
\let\chapterlabel=\relax
\let\sectionlabel=\relax
\newtoks\chapterstyle        \chapterstyle={\Number}
\newtoks\sectionstyle        \sectionstyle={\chapterlabel\Number}
\newskip\chapterskip         \chapterskip=\bigskipamount
\newskip\sectionskip         \sectionskip=\medskipamount
\newskip\headskip            \headskip=8pt plus 3pt minus 3pt
\newdimen\chapterminspace    \chapterminspace=15pc
\newdimen\sectionminspace    \sectionminspace=10pc
\newdimen\referenceminspace  \referenceminspace=25pc
\def\chapterreset{\global\advance\chapternumber by 1
   \ifnum\equanumber<0 \else\global\equanumber=0\fi
   \sectionnumber=0 \makechapterlabel}
\def\makechapterlabel{\let\sectionlabel=\relax
   \xdef\chapterlabel{\the\chapterstyle{\the\chapternumber}.}}
\def\alphabetic#1{\count255='140 \advance\count255 by #1\char\count255}
\def\Alphabetic#1{\count255='100 \advance\count255 by #1\char\count255}
\def\Roman#1{\uppercase\expandafter{\romannumeral #1}}
\def\roman#1{\romannumeral #1}
\def\Number#1{\number #1}
\def\BLANC#1{}
\def\titlestyle#1{\par\begingroup \interlinepenalty=9999
     \leftskip=0.02\hsize plus 0.23\hsize minus 0.02\hsize
     \rightskip=\leftskip \parfillskip=0pt
     \hyphenpenalty=9000 \exhyphenpenalty=9000
     \tolerance=9999 \pretolerance=9000
     \spaceskip=0.333em \xspaceskip=0.5em
     \iftwelv@\fourteenpoint\else\twelvepoint\fi
   \noindent #1\par\endgroup }
\def\spacecheck#1{\dimen@=\pagegoal\advance\dimen@ by -\pagetotal
   \ifdim\dimen@<#1 \ifdim\dimen@>0pt \vfil\break \fi\fi}
\def\TableOfContentEntry#1#2#3{\relax}
\def\chapter#1{\par \penalty-300 \vskip\chapterskip
   \spacecheck\chapterminspace
   \chapterreset \titlestyle{\chapterlabel\ #1}
   \TableOfContentEntry c\chapterlabel{#1}
   \nobreak\vskip\headskip \penalty 30000
   \wlog{\string\chapter\space \chapterlabel} }

\def\section#1{\par \ifnum\the\lastpenalty=30000\else
   \penalty-200\vskip\sectionskip \spacecheck\sectionminspace\fi
   \global\advance\sectionnumber by 1
   \xdef\sectionlabel{\the\sectionstyle\the\sectionnumber}
   \wlog{\string\section\space \sectionlabel}
   \TableOfContentEntry s\sectionlabel{#1}
   \noindent {\caps\enspace\sectionlabel\quad #1}\par
   \nobreak\vskip\headskip \penalty 30000 }
\def\subsection#1{\par
   \ifnum\the\lastpenalty=30000\else \penalty-100\smallskip \fi
   \noindent\undertext{#1}\enspace \vadjust{\penalty5000}}

\def\undertext#1{\vtop{\hbox{#1}\kern 1pt \hrule}}
\def\ack{\par\penalty-100\medskip \spacecheck\sectionminspace
   \line{\fourteenrm\hfil ACKNOWLEDGEMENTS\hfil}\nobreak\vskip\headskip }
\def\APPENDIX#1#2{\par\penalty-300\vskip\chapterskip
   \spacecheck\chapterminspace \chapterreset \xdef\chapterlabel{#1}
   \titlestyle{APPENDIX #2} \nobreak\vskip\headskip \penalty 30000
   \TableOfContentEntry a{#1}{#2}
   \wlog{\string\Appendix\ \chapterlabel} }
\def\Appendix#1{\APPENDIX{#1}{#1}}
\def\appendix{\APPENDIX{A}{}}
\def\unnumberedchapters{\let\makechapterlabel=\relax \let\chapterlabel=\relax
   \sectionstyle={\BLANC}\let\sectionlabel=\relax \sequentialequations }
%
%
%
\def\eqname#1{\relax \ifnum\equanumber<0
     \xdef#1{{\noexpand\rm{\number-\equanumber}}}%
       \global\advance\equanumber by -1
    \else \global\advance\equanumber by 1
      \xdef#1{{\noexpand\rm(\chapterlabel\number\equanumber)}} \fi #1}
\def\eqinsert#1{\noalign{\dimen@=\prevdepth \nointerlineskip
   \setbox0=\hbox to\displaywidth{\hfil #1}
   \vbox to 0pt{\kern 0.5\baselineskip\hbox{$\!\box0\!$}\vss}
   \prevdepth=\dimen@}}
%

%
%
\def\GENITEM#1;#2{\par \hangafter=0 \hangindent=#1
    \Textindent{$ #2 $}\ignorespaces}
\outer\def\newitem#1=#2;{\gdef#1{\GENITEM #2;}}
\newdimen\itemsize                \itemsize=30pt
\newitem\item=1\itemsize;
\newitem\sitem=1.75\itemsize;     
\newitem\ssitem=2.5\itemsize;     
\outer\def\newlist#1=#2&#3&#4;{\toks0={#2}\toks1={#3}%
   \count255=\escapechar \escapechar=-1
   \alloc@0\list\countdef\insc@unt\listcount     \listcount=0
   \edef#1{\par
      \countdef\listcount=\the\allocationnumber
      \advance\listcount by 1
      \hangafter=0 \hangindent=#4
      \Textindent{\the\toks0{\listcount}\the\toks1}}
   \expandafter\expandafter\expandafter
    \edef\c@t#1{begin}{\par
      \countdef\listcount=\the\allocationnumber \listcount=1
      \hangafter=0 \hangindent=#4
      \Textindent{\the\toks0{\listcount}\the\toks1}}
   \expandafter\expandafter\expandafter
    \edef\c@t#1{con}{\par \hangafter=0 \hangindent=#4 \noindent}
   \escapechar=\count255}
\def\c@t#1#2{\csname\string#1#2\endcsname}
\newlist\point=\Number&.&1.0\itemsize;
\newlist\subpoint=(\alphabetic&)&1.75\itemsize;
\newlist\subsubpoint=(\roman&)&2.5\itemsize;
%

%
%
%
%
\newcount\referencecount     \referencecount=0
\newcount\lastrefsbegincount \lastrefsbegincount=0
\newif\ifreferenceopen       \newwrite\referencewrite
\newif\ifrw@trailer
\newdimen\refindent     \refindent=30pt
\def\NPrefmark#1{\attach{\scriptscriptstyle [ #1 ] }}
\let\PRrefmark=\attach
\def\refmark#1{\relax\ifPhysRev\PRrefmark{#1}\else\NPrefmark{#1}\fi}
\def\refend@{\refmark{\number\referencecount}}
\def\refend{\refend@{}\space }
\def\refsend{\refmark{\count255=\referencecount
   \advance\count255 by-\lastrefsbegincount
   \ifcase\count255 \number\referencecount
   \or \number\lastrefsbegincount,\number\referencecount
   \else \number\lastrefsbegincount-\number\referencecount \fi}\space }
\def\refitem#1{\par \hangafter=0 \hangindent=\refindent \Textindent{#1}}
\def\Ref{\rw@trailertrue\REF}
\def\ref{\Ref\?}

\def\REF#1{\r@fstart{#1}%
   \rw@begin{\the\referencecount.}\rw@end}
\def\REFS#1{\r@fstart{#1}%
   \lastrefsbegincount=\referencecount
   \rw@begin{\the\referencecount.}\rw@end}
\def\r@fstart#1{\chardef\rw@write=\referencewrite \let\rw@ending=\refend@
   \ifreferenceopen \else \global\referenceopentrue
   \immediate\openout\referencewrite=referenc.txa
   \toks0={\catcode`\^^M=10}\immediate\write\rw@write{\the\toks0} \fi
   \global\advance\referencecount by 1 \xdef#1{\the\referencecount}}
{\catcode`\^^M=\active %
 \gdef\rw@begin#1{\immediate\write\rw@write{\noexpand\refitem{#1}}%
   \begingroup \catcode`\^^M=\active \let^^M=\relax}%
 \gdef\rw@end#1{\rw@@end #1^^M\rw@terminate \endgroup%
   \ifrw@trailer\rw@ending\global\rw@trailerfalse\fi }%
 \gdef\rw@@end#1^^M{\toks0={#1}\immediate\write\rw@write{\the\toks0}%
   \futurelet\n@xt\rw@test}%
 \gdef\rw@test{\ifx\n@xt\rw@terminate \let\n@xt=\relax%
       \else \let\n@xt=\rw@@end \fi \n@xt}%
}
\let\rw@ending=\relax
\let\rw@terminate=\relax
\let\splitout=\relax
\def\Closeout#1{\toks0={\catcode`\^^M=5}\immediate\write#1{\the\toks0}%
   \immediate\closeout#1}
%
%
\newcount\figurecount     \figurecount=0
\newcount\tablecount      \tablecount=0
\newif\iffigureopen       \newwrite\figurewrite
\newif\iftableopen        \newwrite\tablewrite
\def\FIG#1{\f@gstart{#1}%
   \rw@begin{\the\figurecount)}\rw@end}

\def\Fig{\rw@trailertrue\def\rw@ending{Fig.~\?}\FIG\?}
\def\fig{\rw@trailertrue\def\rw@ending{fig.~\?}\FIG\?}
\def\TABLE#1{\T@Bstart{#1}%
   \rw@begin{\the\tableecount:}\rw@end}
\def\Table{\rw@trailertrue\def\rw@ending{Table~\?}\TABLE\?}
\def\f@gstart#1{\chardef\rw@write=\figurewrite
   \iffigureopen \else \global\figureopentrue
   \immediate\openout\figurewrite=figures.txa
   \toks0={\catcode`\^^M=10}\immediate\write\rw@write{\the\toks0} \fi
   \global\advance\figurecount by 1 \xdef#1{\the\figurecount}}
\def\T@Bstart#1{\chardef\rw@write=\tablewrite
   \iftableopen \else \global\tableopentrue
   \immediate\openout\tablewrite=tables.txa
   \toks0={\catcode`\^^M=10}\immediate\write\rw@write{\the\toks0} \fi
   \global\advance\tablecount by 1 \xdef#1{\the\tablecount}}
%

%
%
%
\def\getfigure#1{\global\advance\figurecount by 1
   \xdef#1{\the\figurecount}\count255=\escapechar \escapechar=-1
   \edef\n@xt{\noexpand\g@tfigure\csname\string#1Body\endcsname}%
   \escapechar=\count255 \n@xt }
\def\g@tfigure#1#2 {\errhelp=\disabledfigures \let#1=\relax
   \errmessage{\string\getfigure\space disabled}}
\newhelp\disabledfigures{ Empty figure of zero size assumed.}
\def\figinsert#1{\midinsert\Tenpoint\medskip
   \count255=\escapechar \escapechar=-1
   \edef\n@xt{\csname\string#1Body\endcsname}
   \escapechar=\count255 \centerline{\n@xt}
   \bigskip\narrower\narrower
   \noindent{\it Figure}~#1.\quad }
%
%
%
\def\masterreset{\global\pagenumber=1 \global\chapternumber=0
   \global\equanumber=0 \global\sectionnumber=0
   \global\referencecount=0 \global\figurecount=0 \global\tablecount=0 }
\def\FRONTPAGE{\ifvoid255\else\vfill\penalty-20000\fi
      \masterreset\global\frontpagetrue
      \global\lastf@@t=0 \global\footsymbolcount=0}

\def\paperstyle{\letterstylefalse\normalspace\papersize}
\def\letterstyle{\letterstyletrue\singlespace\lettersize}
\def\papersize{\hsize=35 truepc\vsize=50 truepc\hoffset=2 truepc
               \skip\footins=\bigskipamount}
\def\lettersize{\hsize=6.5 truein\vsize=8.5 truein\hoffset=0 truein
   \skip\footins=\smallskipamount \multiply\skip\footins by 3 }
\paperstyle   
%
%
\def\MEMO{\letterstyle \letterinfo={\hfil } \let\rule=\memorule
    \FRONTPAGE \memohead }
\let\memohead=\relax

\def\memit@m#1{\smallskip \hangafter=0 \hangindent=1in
      \Textindent{\caps #1}}
\def\subject{\memit@m{Subject:}}
\def\topic{\memit@m{Topic:}}
\def\from{\memit@m{From:}}
\def\to{\relax \ifmmode \rightarrow \else \memit@m{To:}\fi }
\def\memorule{\medskip\hrule height 1pt\bigskip}
\newwrite\labelswrite
\newtoks\rw@toks

\def\addressee#1{\medskip\rightline{\the\date\hskip 30pt} \bigskip
   \ialign to\hsize{\strut ##\hfil\tabskip 0pt plus \hsize \cr #1\crcr}
   \writelabel{#1}\medskip\par\noindent}
\def\rwl@begin#1\cr{\rw@toks={#1\crcr}\relax
   \immediate\write\labelswrite{\the\rw@toks}\futurelet\n@xt\rwl@next}
\def\rwl@next{\ifx\n@xt\rwl@end \let\n@xt=\relax
      \else \let\n@xt=\rwl@begin \fi \n@xt}
\let\rwl@end=\relax
\def\writelabel#1{\immediate\write\labelswrite{\noexpand\labelbegin}
     \rwl@begin #1\cr\rwl@end
     \immediate\write\labelswrite{\noexpand\labelend}}
\newbox\FromLabelBox

\let\labelend=\relax
\def\labelbegin#1\labelend{\setbox0=\vbox{\ialign{##\hfil\cr #1\crcr}}
     \MakeALabel }
\newtoks\FromAddress
\FromAddress={}
\def\MakeFromBox#1{\global\setbox\FromLabelBox=\vbox{\Tenpoint
     \ialign{##\hfil\cr #1\the\FromAddress\crcr}}}
\newdimen\labelwidth        \labelwidth=6in
\def\MakeALabel{\vskip 1pt \hbox{\vrule \vbox{
    \hsize=\labelwidth \hrule\bigskip
    \leftline{\hskip 1\parindent \copy\FromLabelBox}\bigskip
    \centerline{\hfil \box0 } \bigskip \hrule
    }\vrule } \vskip 1pt plus 1fil }
\newskip\signatureskip       \signatureskip=30pt
\def\signed#1{\par \penalty 9000 \medskip \dt@pfalse
  \everycr={\noalign{\ifdt@p\vskip\signatureskip\global\dt@pfalse\fi}}
  \setbox0=\vbox{\singlespace \ialign{\strut ##\hfil\crcr
   \noalign{\global\dt@ptrue}#1\crcr}}
  \line{\hskip 0.5\hsize minus 0.5\hsize \box0\hfil} \medskip }
\newbox\letterb@x
\def\lettertext{\par\unvcopy\letterb@x\par}
\def\multiletter{\setbox\letterb@x=\vbox\bgroup
      \everypar{\vrule height 1\baselineskip depth 0pt width 0pt }
      \singlespace \topskip=\baselineskip }
\def\letterend{\par\egroup}
%
%
%
\newskip\frontpageskip
\newtoks\PubnumI
\newtoks\PubnumII
\newtoks\PubnumIII
\newtoks\PubnumIV
\newtoks\pubtype
\newif\ifp@bblock  \p@bblocktrue
\def\PH@SR@V{\doubl@true \baselineskip=24.1pt plus 0.2pt minus 0.1pt
             \parskip= 3pt plus 2pt minus 1pt }
\def\PHYSREV{\paperstyle\PhysRevtrue\PH@SR@V}
\def\titlepage{\FRONTPAGE\paperstyle\ifPhysRev\PH@SR@V\fi
   \ifp@bblock\p@bblock \else\hrule height\z@ \relax \fi }
\def\nopubblock{\p@bblockfalse}
\def\endpage{\vfil\break}
\frontpageskip=12pt plus .5fil minus 2pt
\pubtype={\tensl Preliminary Version}
\PubnumI={}
\PubnumII={}
\PubnumIII={}
\PubnumIV={}
\def\p@bblock{\begingroup \tabskip=\hsize minus \hsize
   \baselineskip=1.5\ht\strutbox \topspace-2\baselineskip
   \halign to\hsize{\strut ##\hfil\tabskip=0pt\crcr

\the\PubnumI\crcr\the\PubnumII\crcr\the\PubnumIII\crcr\the\PubnumIV\crcr
        \the\date\crcr\the\pubtype\crcr}\endgroup}
\def\title#1{\vskip\frontpageskip \titlestyle{#1} \vskip\headskip }
\def\author#1{\vskip\frontpageskip\titlestyle{\twelvecp #1}\nobreak}

\def\address#1{\par\kern 5pt\titlestyle{\twelvepoint\it #1}}
\def\andaddress{\par\kern 5pt \centerline{\sl and} \address}

\def\abstract{\par\dimen@=\prevdepth \hrule height\z@ \prevdepth=\dimen@
   \vskip\frontpageskip\centerline{\fourteenrm ABSTRACT}\vskip\headskip }

%
%
%

\def\\{\relax \ifmmode \backslash \else {\tt\char`\\}\fi }
\def\sequentialequations{\relax\if\equanumber<0\else\global\equanumber=-1\fi}

\def\journal#1&#2(#3){\unskip, \sl #1\unskip~\bf\ignorespaces #2\rm (19#3),}

\def\topspace{\hrule height 0pt depth 0pt \vskip}

\def\Buildrel#1\under#2{\mathrel{\mathop{#2}\limits_{#1}}}
\def\becomes#1{\mathchoice{\becomes@\scriptstyle{#1}}{\becomes@\scriptstyle
   {#1}}{\becomes@\scriptscriptstyle{#1}}{\becomes@\scriptscriptstyle{#1}}}
\def\becomes@#1#2{\mathrel{\setbox0=\hbox{$\m@th #1{\,#2\,}$}%
    \mathop{\hbox to \wd0 {\rightarrowfill}}\limits_{#2}}}

\let\int=\intop         
\def\lsim{\mathrel{\mathpalette\@versim<}}
\def\gsim{\mathrel{\mathpalette\@versim>}}
\def\@versim#1#2{\vcenter{\offinterlineskip
    \ialign{$\m@th#1\hfil##\hfil$\crcr#2\crcr\sim\crcr } }}
\def\big#1{{\hbox{$\left#1\vbox to 0.85\b@gheight{}\right.\n@space$}}}
\def\Big#1{{\hbox{$\left#1\vbox to 1.15\b@gheight{}\right.\n@space$}}}
\def\bigg#1{{\hbox{$\left#1\vbox to 1.45\b@gheight{}\right.\n@space$}}}
\def\Bigg#1{{\hbox{$\left#1\vbox to 1.75\b@gheight{}\right.\n@space$}}}
%
%
%
\let\sec@nt=\sec
\def\sec{\relax\ifmmode\let\n@xt=\sec@nt\else\let\n@xt\section\fi\n@xt}
\def\obsolete#1{\message{Macro \string #1 is obsolete.}}
\def\firstsec#1{\obsolete\firstsec \section{#1}}
\def\firstsubsec#1{\obsolete\firstsubsec \subsection{#1}}
\def\thispage#1{\obsolete\thispage \global\pagenumber=#1\frontpagefalse}
\def\thischapter#1{\obsolete\thischapter \global\chapternumber=#1}
\def\REFSCON{\obsolete\REFSCON\REF}
\def\splitout{\obsolete\splitout\relax}
\def\prop{\obsolete\prop \propto }
\def\nextequation#1{\obsolete\nextequation \global\equanumber=#1
   \ifnum\the\equanumber>0 \global\advance\equanumber by 1 \fi}
\def\BOXITEM{\afterassigment\B@XITEM\setbox0=}
\def\B@XITEM{\par\hangindent\wd0 \noindent\box0 }
\def\phyzzx{PHY\setbox0=\hbox{Z}\copy0 \kern-0.5\wd0 \box0 X}
%
%
\everyjob{\xdef\today{\monthname\ \number\day, \number\year}}
\PubnumI={}
\PubnumII={}
\PubnumIII={}
\PubnumIV={}
\pubtype={}
\def\memohead{\line{\fourteenrm Texas A\&M University:\ \twelverm
      Department of Physics.\hfil\twelveit \the\date}\medskip}
 at 7 truept
 at 10. truept
 at 12.00 truept
 at 7 truept
 at 10. truept
 at 12.00 truept

%
\FromAddress={\crcr Center for Theoretical Physics\cr Texas A\&M University\cr
    College Station, Texas 77843-4242\crcr}
%
\Letterinfo={\ninerm \hfil Phone: (409) 845--7773\qquad
                        Bitnet: LOPEZ@TAMPHYS\qquad
                      Fax: (409) 845--2590\hfil }
\edef\memorule{\medskip\hrule\kern 2pt\hrule \noindent
      \llap{\vbox to 0pt{ \vskip 1in\normalbaselines \tabskip=0pt plus 1fil
        \halign to 0.99in{\seventeenrm\hfil ##\hfil\cr
          M\cr E\cr M\cr O\cr R\cr A\cr N\cr D\cr U\cr M\cr}
        \vss }}\par\medskip}
\newread\figureread
\def\g@tfigure#1#2 {\openin\figureread #2.fig \ifeof\figureread
    \errmessage{No such file: #2.fig}\xdef#1{\hbox{}}\else
    \read\figureread to\y@p \read\figureread to\y@p
    \read\figureread to\x@p \read\figureread to\y@m
    \read\figureread to\x@m \closein\figureread
    \xdef#1{\hbox{\kern-\x@m truein \vbox{\kern-\y@m truein
      \hbox to \x@p truein{\vbox to \y@p truein{
        \special{include(#2.fig relative)}\vss }\hss }}}}\fi }
\catcode`\@=12 
\message{ by V.K., G.S.}

\sequentialequations
\hfuzz=100pt
\catcode`@=11
\def\lrarrowfill{$\m@th \mathord\leftarrow\mkern-6mu\cleaders
           \hbox{$\mkern-2mu \mathord- \mkern-2mu$}\hfill
           \mkern-6mu\mathord\rightarrow$}
\catcode`\^^@=9
\def\overlrarrow#1{\vbox{\ialign{##\crcr
     \lrarrowfill\crcr\noalign{\kern-1pt\nointerlineskip}
     $\hfil\displaystyle{#1}\hfil$\crcr}}}

\def\rl{\rightline}
\def\ll{\leftline}

\def\etal{{\it et. al.}}

\def\t1{{\tilde 1}}

\def\HP{Heath Pois}
\def\TJW{Thomas J. Weiler}
\def\TC{Tzu Chiang Yuan}

\def\GeV{\,{\rm GeV}}

\def\NPB#1#2#3{Nucl. Phys. {\bf B#1}, #2 (19#3)}
\def\PLB#1#2#3{Phys. Lett. {\bf B#1}, #2 (19#3)}
\def\PRD#1#2#3{Phys. Rev. {\bf  D#1}, #2 (19#3)}
\def\PR#1#2#3{Phys. Rev. {\bf  #1}, #2 (19#3)}
\def\PRL#1#2#3{Phys. Rev. Lett. {\bf#1}, #2 (19#3)}

\REF\chanowitz{M. Chanowitz and M. Gaillard, \NPB{261}{379}{85}.}
\REF\hunter{See for example, J. F. Gunion, H. E. Haber, G. L. Kane, and
S. Dawson, {\it The Higgs Hunters' Guide}, Addison-Wesley,
Redwood City, CA (1990) and references therein.}
\REF\lep{M. Davier, Talk given at LP-HEP91, July, 1991.
The new ALEPH bound is $M_{H_{SM}}> 54\GeV$.
D. Decamp {\it et al.,} (ALEPH Collaboration),
Phys. Lett. {\bf B246}, 306
(1990); P. Abreu {\it et al.,} (DELPHI Collaboration), Nucl. Phys. {\bf B342},
1 (1990); B. Adeva {\it et al}., (L3 Collaboration), Phys. Lett. {\bf B257},
452 (1991); M. Z. Akrawy {\it et al}., Phys. Lett. {\bf B253}, 511 (1991).}
\REF\cdf{F. Abe {\it et al.} (CDF Collaboration),
Phys. Rev. Lett. {\bf 64}, 147 (1990).}
\REF\mtop{For recent analysis see
P. Langacker and M. Luo, \PRD{44}{817}{91} and references therein.}
\REF\veltman{M. Veltman, Acta Phys. Pol. {\bf B8}, 475 (1977).}
\REF\rizzo{J. Ellis, Talk Given at the LP-HEP CERN Conference, July, 1991;
T. G. Rizzo, Mod. Phys. Lett. {\bf A6}, 2947 (1991).
See also W. Hollik, Lectures at the $21^{st}$ G.I.F.T., Santander, Spain,
June 1990, and Proc. $13^{th}$ Warsaw Symp. on Elem. Particles, Kazimierz,
Poland, 1990.}
\REF\susyhm{H. E. Haber and R. Hempfling, Phys. Rev. Lett. {\bf 66}, 1815
(1991); J. Ellis, G. Ridolfi, and F. Zwirner, Phys. Lett. {\bf B257}, 83
(1991) and {\bf B 262}, 477 (1991);
R. Barbieri, M. Frigeni, and F. Caravaglios, Phys. Lett. {\bf B258},
167 (1991); J. L. Lopez and D. V. Nanopoulos, Phys. Lett.
{\bf B266}, 397 (1991);
Y. Okada, M. Yamaguchi, and T. Yanagida, Prog. Theor. Phys.
{\bf 85} 1 (1991). See also P. H. Chankowski, S. Pokorski, and J. Rosiek,
Phys. Lett. {\bf B274}, 191 (1992).}
\REF\GKW{J. F. Gunion, G. L. Kane, and J. Wudka,
\NPB{299}{231}{88}.}
\REF\KM{W.--Y. Keung and W. J. Marciano, \PRD{30}{248}{84}.}
\REF\BARGER{V. Barger, G. Bhattacharya, T. Han, and B. A. Kniehl,
\PRD{43}{779}{91}.}
\REF\EEEE{C. N. Yang, Phys. Rev. {\bf 77}, 722 (1950); N. M. Kroll
and W. Wada, Phys. Rev. {\bf 98}, 1355 (1955).}
\REF\history{N. P. Samios, R. Plano, A. Prodell, M. Schwartz, and J.
Steinberger, Phys. Rev. Lett. {\bf 3}, 524 (1959); Phys. Rev. {\bf 126}, 1844
(1961). The non-observation of the process $\pi^- + d \rightarrow 2n+ \pi^0$,
expected to be highly suppressed if the $\pi^0$ has odd parity, provided
earlier circumstantial evidence for the $\pi^0$'s parity: W. Chinowsky and
J. Steinberger, Phys. Rev. {\bf 100}, 1476 (1955).}
\REF\farhi{For example, in technicolor models, the technifermions
condense to generate
a Higgs mechanism, while the fermions acquire their masses from a separate
``extended'' technicolor sector. See E. Farhi and L. Susskind,
Phys. Rep. {\bf 74C}, 2777 (1981).}
\REF\HKS{A two-doublet model with one of the doublets
yielding `fermiophobic' Higgs particles has been discussed some time ago by
H. E. Haber, G. L. Kane, and T. Sterling, \NPB{161}{553}{79},
App.B.}
\REF\weiler{T. J. Weiler, VIII Vanderbilt International Conference on
High Energy Physics, edited by
J. Brau and R. S. Panvini, 1987, World Scientific
Publications.}
\REF\yang{C. N. Yang, Phys. Rev. {\bf 77}, 242 (1950).}
\REF\dkr{M. J. Duncan, G. L. Kane, and W. W. Repko, Nucl. Phys.
{\bf B272}, 517 (1986); J. R. Del'Aquila and C. A. Nelson,
\PRD{33}{80}{86};
{\it ibid}, 93 (1986); C. A. Nelson, \PRD{37}{1220}{88}; T. Matsura and J. J.
van der Bij, Z. Phys. {\bf C51},259 (1991).}
\REF\lastref{Last reference in [18].}
\REF\duncan{M. J. Duncan, Phys. Lett. {\bf B179}, 393 (1986), and first
reference of [18].}
\REF\inverse{The inverse process, $t$-quark $\rightarrow$
Higgs + light fermion + ($W$ or $W^{\ast}$) has been
considered by T. G. Rizzo, Phys. Rev. {\bf D35}, 1067 (1987);
V. Barger and W.--Y. Keung, \PLB{202}{393}{88}.}
\REF\ttbar{W. A. Bardeen, C. T. Hill, and M. Lindner, \PR{41}{1647}{90} and
references therein. Bubble sums yield a weakly bound top condensate with
$M_H\sim 2m_t$; renormalization group improved sums yield a more strongly
bound result of $M_H\sim 1.1m_t$. }
\REF\div{P. Osland and T. T. Wu, in CERN preprint TH-6567 (1992), use
cancellation of certain divergences to fix $M_H \sim 190\;GeV$ and $m_t
\sim 120\;GeV.$}
\REF\paige{F. E. Paige, Brookhaven preprint \#45828, Jan. (1991).}
\REF\marcpaige{W. J. Marciano and F. E. Paige, \PRL{66}{2433}{91};
J. F. Gunion, \PLB{261}{51}{91}.}
\REF\wlong{In the Higgs decay chain, both the Higgs and top decay to $W$'s.
For $M^2_H \gg M^2_W$, the ratio of Higgs-produced longitudinal
$W$'s to transverse
$W$'s is $M_H^4 / 8 M_W^4$, while for top-produced $W$'s, the same ratio
is $m_t^2 / 2 M_W^2+{\cal O}(m^2_b/M^2_W)$.
These ratios are easy to understand when the equivalence theorem is invoked.}
\REF\bspm{V. Barger, A. L. Stange, and R. J. N. Phillips,
\PRD{44}{1987}{91};
{\it ibid}, \PRD{45}{1484}{92}.}
\REF\pq{R. Peccei and H. Quinn, \PRD{16}{1791}{77}.}
\REF\nonhiggs{Most recently, charged singlet scalars have been invented to
generate a large transition magnetic moments for the light neutrinos; see
S. M. Barr and E. M. Friere, \PRD{43}{2989}{91} and references therein.}
\REF\simmons{E. H. Simmons, Nucl. Phys. {\bf B312}, 253 (1989); {\it ibid},
{\bf B324}, 315 (1989); A. Kagan and S. Samuel, \PLB{252}{605}{90};
{\it ibid}, Int. J. Mod. Phys. {\bf A7}, 1123 (1992).}
\REF\KWMT{N. M. Kroll and W. Wada, ref. [\EEEE];
T. Miyazaki and E. Takasugi, Phys. Rev. {\bf D8}, 2051 (1973).}

\rl{VAND--TH--93--4}

\nopagenumbers

\title{\bf \centerline{HIGGS-BOSON DECAY TO FOUR FERMIONS} \hfil\break
INCLUDING A SINGLE TOP QUARK BELOW $t \bar t$ THRESHOLD}
\author{\HP$^{1,2}$, \TJW$^3$, and \TC$^{4*}$}
\address{$^1$ Center For Theoretical Physics, Department of Physics, \break
Texas A\&M University, College Station, TX 77843-4242}
\address{$^2$ Astroparticle Physics Group, Houston Advanced Research Center
(HARC)\break
The Woodlands, TX 77381, USA}
\address{$^3$ Department of Physics and Astronomy, \break
Vanderbilt University, Nashville,
TN 37235}
\address{$^4$ Department of Physics and Astronomy, \break
Northwestern University, Evanston, IL 60208}

\vfill

\ll{Email address: pois@tamphys, weilertj@vuctrvax, nuhep::yuant}

\ll{* Present address: Department of Physics, University of California at
Davis, CA 95616}

\endpage

\abstract

The rare decay modes Higgs $\rightarrow$ four light fermions, and
Higgs $\rightarrow$ single top-quark + three light fermions for $m_t<M_H<2m_t$,
are presented, and phenomenologically interpreted. The angular correlation
between fermion planes is presented as a test of the spin and intrinsic
parity of the Higgs particle. In Higgs decay to single top, two tree-level
graphs contribute in the standard model (SM);
one couples the Higgs to $W^+W^-(\sim gM_W)$, and
one to $t\bar t(\sim g_{top\;yukawa}=m_t/246\GeV)$.
The large Yukawa coupling for $m_t>100\GeV$ makes the second amplitude
competitive or dominant for most $M_H,m_t$ values. Thus the Higgs decay
rate to single top directly probes the SM universal mechanism generating
both gauge boson and fermion masses, and offers a means to infer the
Higgs-$t \bar t$ Yukawa coupling when $H\rightarrow t \bar t$ is
kinematically disallowed.  We find that the modes $pp\rightarrow
Xt\bar t(H\rightarrow t\bar b W^{(*)})$ at the SSC, and
$e^+ e^-\rightarrow Z\,or\,\nu\bar{\nu} + (H\rightarrow t\bar b W^{(*)})$
at future high energy, high luminosity
colliders, may be measureable if
$2m_t$ is not too far above $M_H$. We classify non-standard Higgses as
gaugeo-phobic, fermio-phobic or fermio-philic,
and discuss the Higgs$\rightarrow$ single top rates for these classes.

\endpage

\pagenumbers
\normalspace
\pagenumber1
\centerline{1. INTRODUCTION}

The existence of a Higgs scalar particle, or the onset of new physics at a few
TeV or less, is a virtually
guaranteed happenstance in particle physics [\chanowitz].
Either possibility provides a
foundation for the electroweak unification  and the generation of masses.  Of
the two, the
simpler possibility is the existence of the Higgs particle.  Accordingly, a
tremendous amount of energy has been, and continues to be, devoted to
theoretical and experimental searches for signatures of the Higgs
boson [\hunter]. The four experiments at LEP have recently
placed a lower bound of $M_H \sim 57\GeV @95\% CL$ for
a Standard Model (SM) Higgs [\lep].
With regard to the top quark mass, a direct SM lower bound
of $m_t \sim 89$ GeV has also been obtained from the CDF experiment [\cdf].

Indirect SM upper bounds for $m_t$ and $M_H$ can be predicted
by the theory based on  quantum loop
phenomenology. It is well known that heavy top quark
loop corrections to certain low energy and electroweak (EW) observables
(for example, the $\rho$
parameter) are proportional to $m_t^2$
and thus the quantum effects are quite sensitive to $m_t$.
The SM consistency of all the
low energy experimental data requires $m_t<182 \;\GeV @ 95\% CL$,
with a central value of $m_t=125\pm30\GeV$ [\mtop]. On the
other hand, the mass dependence of  a heavy Higgs loop correction
varies as ${\rm ln} M_H^2$ in the SM. This is the famous
one-loop ``screening rule" first recognized by Veltman [\veltman].
Since the dependence of quantum loop effects on
the heavy Higgs is only logarithmic, low energy observables are not
very sensitive to $M_H$.
However, recent analysis [\rizzo] indicates that a weak upper bound
($M_H \sim 300-500 \; \GeV$) for the SM
Higgs can be deduced from current low energy experimental data which is
suggestive of a `light' Higgs.
In models with a broken supersymmetry, the tree level mass of the lightest
Higgs generally
lies below the $Z$ mass [\hunter]. However,
the large radiative corrections to the Higgs boson masses due to a heavy
top quark [\susyhm] may alter this situation significantly.

Finding a signature for
an `intermediate mass' Higgs ($M_W<M_H<2M_Z$) is particularly problematical.
It appears that
rare decay modes of the Higgs such as $H\rightarrow Z \gamma,\gamma \gamma,
\tau^+\tau^-,b\bar b, \Theta \gamma$
offer the greatest promise for providing an
experimental signature of the particle's existence [\GKW].  The decay of a
Higgs with mass above the two $W$ threshold is dominantly to two $W$'s, and
so one might believe that a Higgs with mass not too far
below twice the $W$ mass may have a
significant branching fraction to decay through  $WW^*$ or $W^*W^*$
intermediate states ($W^*$ denotes a virtual $W$).

Marciano and Keung [\KM] have shown that for a standard model Higgs with mass
below but near the two $W$ threshold, the branching ratio for the decay
$H \rightarrow WW^{\ast}$ becomes significant;
in fact, this decay mode
dominates all others if the decay to $t\bar t$ is kinematically forbidden
(i.e. $M_H<2m_t$). Barger \etal [\BARGER] have considered the
$W^*W^* \; \rightarrow$ four fermions decay mode in the massless
fermion limit, and have made a thorough
study of the SSC signal and backgrounds.
Their conclusion is that only if $m_t,M_H>150\GeV$ will the decay signal be
seen

above the $t\bar t\rightarrow W\bar b W b$ and $W$ continuum background
channels.

Our results here are complementary to the work in [\BARGER], and extend that
work by including a single massive top quark in the final state.
As in [\BARGER], we allow
both $W$'s to be virtual, and let phase space
optimize the sharing of virtuality
between the two $W$'s.
Our exact results for Higgs $\rightarrow$ four fermions
allow us to exhibit the correlations
among final fermion energies and angles that test the presumably
scalar nature of the
parent Higgs particle. As an example of final state correlations, we display
the dependence of the decay on the angle between the decay planes defined for
each vector boson.  The analogous azimuthal angular dependence
for the process pion $\rightarrow e^+e^-e^+e^-$
was calculated over thirty years ago [\EEEE],
and ultimately provided the signature distinguishing between a scalar and a
pseudoscalar `pion' [\history].

In Sec.2 we present our
result for the massless fermion limit (which agrees with the
result in [\BARGER]), and we discuss a class of nonstandard
Higgses which do not couple directly to fermions. Such `fermiophobic' Higgs
decay through loops to two fermions, or through two gauge bosons to
four fermions.  The `fermiophobic' Higgs branching ratio to four
fermions is large, even if mass kinematics require both gauge bosons
to be virtual; a calculation of
this branching ratio requires the exact formulae presented in this paper.
In Sec.3 we define the asymmetry parameters relevant
to the angular correlations between the fermion planes, and explore the
$M_H$ and virtual gauge-boson mass dependence of these parameters.

In Sec.4 we exhibit the rate for the tree level decay of the Higgs
to a single heavy quark (e.g. top) plus three light fermions.  Since the Higgs
particle has a direct Yukawa coupling to fermions which scales with the
fermion mass, this latter calculation includes a second graph with a
potentially large contribution to the amplitude, and therefore
provides an important contrast to the massless fermion case.
We find that due to the large Yukawa coupling for large $m_t$,
the branching ratio to single top (e.g. $H \rightarrow t\bar b
s \bar c$)
can be almost competitive with the dominant massless fermion decay mode
$H\rightarrow WW\rightarrow 4f$, and is certainly competitive with the
rare $H\rightarrow
\gamma \gamma , Z \gamma, b \bar b, \tau^+\tau^-$ decay modes for a range
of $M_H,m_t$ values. In Sec. 5, we present a brief discussion of the
signature and backgrounds for the $H\rightarrow t b W^*$ mode. We
argue that the signal may be detectible at hadron colliders for Higgses
produced in association with a $t\bar t$ pair, and at future high energy,
high luminosity $e^+e^-$ colliders. Interesting non-standard
Higgs models are classified in Sec. 6, and the branching ratios to
single top of these baroque Higgses are discussed. We conclude in Sec. 7.
Detailed formulae for the decay of a Higgs boson to four fermions,
including the massive heavy top, are collected in an appendix.

\vskip 1in

\centerline{2. HIGGS DECAY TO FOUR MASSLESS FERMIONS}

For $H\rightarrow WW,ZZ\rightarrow 4f$ (where the $W$'s and $Z$'s may be
real or virtual), the relevant Feynman
diagram when $m_f=0$ is shown in Fig.1(a);
Fig.1(b) obviously does not contribute in
the massless limit to the SM Higgs due to the vanishing Yukawa coupling.
(See however Sec. 6 for non--standard Higgses). The result
for the Higgs width to four fermions is presented in the appendix.
The differential width is symmetric in the invariant masses $Q_1^2$
and $Q_2^2$ of the $W$ or $Z$ gauge bosons, and
we show the gauge boson mass spectrum in Fig. 2 for the process
$H\rightarrow WW\rightarrow u\bar d s \bar c + \bar u d \bar s c$.
One can observe that once $M_H>M_W$, a
bimodal distribution results.
The twin peaks correspond to one or
the other $W$ being on-shell; for invariant mass values $M_{inv}$
between the peaks,
kinematics requires both $W$'s to be off-shell.  It is clear that the relative
contribution with both $W$'s off-shell is small for an intermediate mass
Higgs.  This observation is borne out in a comparison between
the exact four fermion width, and the one gauge boson on-shell (OGBOS)
approximate width [\KM],
which is obtained by replacing one of the Breit-Wigner factors
$((Q^2-M_V^2)^2+M_V^2\Gamma_V^2)^{-1}$ in
the decay width with the narrow width approximate (NWA) form
$\pi \delta (Q^2-M_V^2)/M_V\Gamma_V$.
In agreement with [\BARGER], we
find that putting one $W$ on-shell is a good approximation to the
total decay rate for $M_H>100\GeV$.  For Higgs decay through
two $Z$'s, we find that putting one $Z$ on-shell approximates well
the true decay rate if $M_H>115 \GeV$.
In conclusion, we find that the rate for a SM Higgs
particle to decay to four fermions is a percent of that for decay
to $b\bar b$, for $M_H \sim 100$ GeV, and falls rapidly
for lighter Higgs masses; for $M_H > 100$ GeV, the OGBOS approximation
gives an accurate rate.

A bit of caution is required when applying the NWA to
the Breit-Wigner integrals of
Eqn.(A.21). If one Breit-Wigner is replaced by the NWA delta function,
a multiplicative factor of two
must be introduced to account for the probability for either $W$
to be on-shell.
Above the two $W$ threshold, this factor of two must be removed since one
$W$
on-shell no longer precludes the possibility of the second $W$ also being
on-shell.  If
Eqn.(A.21) is used without approximation, there
is no need to concern oneself with this extra counting.

The four-fermion mode becomes particularly interesting in models where
the two-fermion modes are suppressed at tree level.
Recall that in the SM with three families of quarks and leptons, the
Higgs doublet plays a double role of generating masses for both the gauge
sector and fermion sector. In fact, there is a third role:
the vacuum expectation
value (VEV) of the Higgs field breaks not only the $SU(2) \times U(1)$
gauge symmetry
but also the flavor symmetry $[U(3)]^5$ (the three copies of the five fields
$l_L,q_L,e_R,u_R,$ and $d_R$).
The Higgs mechanism, originally employed to
give nonzero masses to the gauge bosons while maintaining a renormalizable
theory, may not be the source of the fermion masses or the flavor
symmetry breaking.  The independent
generation of
fermion masses and gauge bosons masses is conceptually possible [\farhi].
A suppression of the Higgs couplings to fermions is easily enforced by
introducing a discrete $\phi \rightarrow -\phi$ symmetry;
the symmetry forbids Yukawa couplings [\HKS].  We call
the resulting Higgs fields `fermiophobic', for although the fields couple to
the gauge bosons at tree level, their coupling to fermions occurs only through
loops and scalar field mixing (The loop and mixing-induced couplings are
allowed since $<\phi> \ne  0$ breaks the discrete symmetry).

It is a logical possibility then
that the sole Higgs particle introduced in the standard model is itself
`fermiophobic'.  If so, the rate for a Higgs of intermediate mass
to decay to four fermions via two virtual gauge bosons (Fig.1(a))
competes favorably with all other decays,
namely Higgs $\rightarrow$ two
fermions through scalar mixing and/or loop graphs,
Higgs $\rightarrow \gamma \gamma$, or
$\gamma Z$ through a $W$ or charged Higgs loop, and Higgs $\rightarrow gg$
through one-loop and mixing, or through two loops.
In fact, the decay rate to four fermions exceeds the rate to
$\gamma \gamma$ or  $\gamma Z$ if the Higgs mass exceeds 75 GeV.
Branching ratios for a fermiophobic Higgs may be inferred from the SM
branching ratios presented in Fig.2 of ref. [\BARGER].
If an intermediate mass Higgs
is discovered, a comparison of its two fermion decay rate to
its four fermion or two photon rate will immediately provide important
information on the mechanism(s) of mass generation. [\weiler]

\vskip 1in

\centerline{3. ANGULAR CORRELATIONS BETWEEN THE FERMION PLANES}

Next we turn to the correlations among the final state
of four massless fermions implied by the
matrix element of Eqn.(A.4). The reader
may recall that parity conservation
implies that the two photons resulting from a scalar particle decay have
parallel polarization directions, while the photons from a decaying
pseudoscalar particle have perpendicular polarizations [\yang].
A similar result holds
for decay of scalar and pseudoscalar bosons to two $Z$'s or two $W$'s with
momenta $Q_1$ and $Q_2$ and polarization vectors
$\epsilon_{1}^{\mu} (\lambda) \; {\rm and} \;  \epsilon_{2}^{\nu}
(\lambda^{\prime})$. The matrix element is linear in the two
polarization vectors.

In the
case of even-parity Higgs field surviving spontaneous symmetry
breaking, the decay proceeds at the tree level with
the two polarization vectors contracted
$\epsilon_1(\lambda)\cdot  \epsilon_2(\lambda^{\prime})$,
which tells us that the two
gauge bosons emerge with the same linear polarization.  In a helicity basis,
the two gauge bosons emerge with the same helicity.  For massive gauge bosons,
there are of course three physical polarizations,
and a fourth `scalar' polarization for off-shell vector bosons which
vanishes when coupled to a conserved current.  In the case of the odd-parity
neutral pseudoscalar decay, CP invariance of the gauge boson and scalar
boson sectors is sufficient to ensure no tree level coupling [\hunter]; the
process is loop induced with $\phi F {\tilde F}$ being the lowest dimensional
operator, just as in the $\pi^0 \rightarrow \gamma \gamma$ case. The
Lorentz invariant resulting from this operator,
linear in the two polarization vectors, is
$\epsilon_{\mu\nu\alpha\beta} \epsilon^{\mu}_1(\lambda)
\epsilon_2^{\nu}(\lambda^{\prime})
Q^{\alpha}_1Q^{\beta}_2$ (times a coefficient down by
$\alpha$ compared to the tree level scalar case).
In the pseudoscalar particle
rest frame this invariant is proportional to
${\vec \epsilon}_1(\lambda) \times
{\vec \epsilon}_2(\lambda^{\prime})\cdot
({\vec Q}_1 - {\vec Q}_2)$,
which tells us
that only transverse polarizations, or equivalently, left and right helicities,
are produced.  Furthermore, the linear polarization vectors are perpendicular
to each other.  In a helicity basis, the two helicities are again identical.
We expect the final decay planes to remember the
vector boson polarizations, just as in electromagnetic decay of scalar or
pseudoscalar particles to two $e^+e^-$ pairs [\EEEE].  Since a scalar parent
particle gives a Lorentz invariant combination of polarization
vectors differing from that of a pseudoscalar parent, we expect the
angular orientation of the final fermion decay
planes to be an indicator of the parity of the parent particle.

So far, our discussion of polarization has concerned itself only with vector
polarization.  In fact, since the polarization of the gauge bosons cannot be
directly measured, a true description of the gauge boson polarization
requires a
density matrix; the vector polarization is just the diagonal part of the
general density matrix.
The density matrix for an off-shell vector boson is a $4 \times 4$ matrix.
Since our calculation couples the gauge boson to a conserved final state
current (in the massless fermion approximation), only the `physical'
$3 \times 3$ sector contributes.
When either gauge boson
is put on-shell and treated as an observable particle, as with the OGBOS
approximation, the correlations contained in the off-diagonal elements
are lost! Of course, all of the
correlations implied by the density matrix formalism are contained in the
exact result of Eqns.(A.17-21), which we use.

Let us define the azimuthal asymmetry parameters $\alpha_1$ and $\alpha_2$
according to the following formulae:
$$
{d\Gamma \over d\phi} = {\Gamma \over {2\pi}}
\bigg[ 1+\alpha_1(V)\cos\phi+\alpha_2(V)\cos 2\phi \bigg] \eqno(1)
$$
$\phi$ is the angle between the two planes, where each plane is defined by the
two fermion momentum vectors resulting from the decay of a virtual or real
gauge boson, as measured in the four fermion CMS (i.e. the Higgs boson rest
frame).  $\phi$ is invariant under boosts in the direction of the momentum
of either gauge boson.
$V$ stands for either the $WW$ decay mode, or the
$ZZ$ mode.
The expressions for $\alpha_1(V)$ and $\alpha_2(V)$ for scalar Higgs decay
are given in Eqns.(A.17,A.18). $\alpha_1(V)$ is a parity violating asymmetry
and $\alpha_2(V)$ is parity conserving. We show in Fig.3 the
asymmetry parameters as a function of the Higgs mass $M_H$,
for the decays $H\rightarrow ZZ\rightarrow d \bar d s \bar s$, and
$H\rightarrow WW\rightarrow {u \bar d}s \bar c+ \bar u d\bar s c$.
For convenience, we have chosen
final states such that identical fermions, requiring antisymmetrization, do
not occur, and such that interference between the two-$W$ graph and the
two-$Z$ graph does not occur.  It is seen that the asymmetries are largest for
the intermediate mass Higgs, and peak near the threshold values
$M_H=2M_W,2M_Z$ for
$\alpha_{1,2}(W),\alpha_{1,2}(Z)$ respectively.  At threshold the
two $V$'s are nearly on--shell and at rest, so the three nonzero helicity
amplitudes are equal and the off--diagonal density matrix elements are
maximized.
In particular, the parity violating asymmetry  $\alpha_1(V)$
can be of ${\cal O}$(1) for the intermediate mass Higgs,
and may provide a clean
consistency check on the scalar nature of the decaying Higgs particle. In
comparison, from Fig.3 we see that the parity conserving
asymmetry is  reduced by an order of magnitude or more.

Unfortunately, the large asymmetry $\alpha_1$ is difficult to determine
experimentally. The azimuthal angle may be written as $\cos \phi = \hat
n _1 \cdot \hat n _2$, where $\hat n _{1,2}$ are the normals to the
two $WW$ or $ZZ$ decay planes. To determine these normals, an experimenter
must have charge and possibly flavour identification of the emitted
leptons and/or jets. An $f \leftrightarrow \bar f$ ambiguity results in
a $\hat n \leftrightarrow - \hat n$ or $ \phi \leftrightarrow
\pi - \phi$ ambiguity, and washes out the $\cos \phi$ asymmetry $\alpha_1$.
On the other hand, the $\cos 2\phi$ term is even under this ambiguity,
but $\alpha_2$ is small except near $VV$ threshold.

The asymmetries have been calculated before for on-shell
$W$'s and $Z$'s [\dkr], and for the QCD background process $gg,q\bar q
\rightarrow ZZ\rightarrow 4f$ [\lastref]. Applying the NWA to our more
general results, Eqns. (A.17-A.18), or referring to the literature, one
finds simple expressions:
$$
\alpha_1({\rm on-shell} \; V's) =
{9\pi^2\over 16}\left( {C_R-C_L \over C_R+C_L} \right)
{M_V^2(M_H^2-2M_V^2) \over
{\left[ (M_H^2-2M_V^2)^2 + 8M_V^4 \right]}} \quad , \eqno(2)
$$
$$
\alpha_2({\rm on-shell} \; V's) = {2M_V^4 \over
{\left[ (M_H^2-2M_V^2)^2 + 8M_V^4 \right]}} \quad , \; V=W,Z \quad . \eqno(3)
$$
At threshold $\alpha_2(V) = 1/6$ (evident in our figure);
above threshold $\alpha_2(W)$ and $\alpha_2(Z)$ fall rapidly
(asymptotically, like $(M_V/M_H)^4$)
to less than two percent at
$M_H \geq $ 300 GeV. What is new in our calculations are the below
threshold results: $\alpha_2(V)$ falls as $M_H$ moves below the $VV$
threshold, but slowly. When kinematics require one (two) $V$ off-shell,
the asymmetry is $\simeq 10\% (7\%)$. For $\alpha_1(V)$, the threshold
value is ${3\pi^2\over 32}\left( {C_R-C_L \over C_R+C_L} \right)$.
$C_R$ and $C_L$ are given in the appendix. For the $V=W$ case, (where
parity violation is maximal, and $C_L=1,C_R=0$) $\alpha_1(W)={-3\pi^2\over 32}
=-0.925$, nearly providing a zero in ${d\Gamma \over d\phi}$ at
$\cos \phi=1$. Extremization of $\alpha_1(V)$ with respect to $M_H$
reveals that $|\alpha_1(V)|$ achieves its maximum value a bit above
threshold, at $M_H=2.35M_V$, i.e. at $188(214)\; \GeV$ for the $W(Z)$ boson.
This slight displacement is also evident in our figure. The above
threshold fall-off is not so rapid as for $\alpha_2$, going asymptotically
like $(M_V/M_H)^2$. The below threshold fall-off for $\alpha_1$ is also
weaker than for $\alpha_2$, dropping by $15\% (30\%)$
when one (two) $V$'s are kinematically moved off-shell.

The asymmetries arise from interference among the possible $H\rightarrow
V^{(*)}V^{(*)}$ helicity amplitudes. In terms of helicity amplitudes, the
dependence of $\alpha_1$ and $\alpha_2$ on $M_H$ is qualitatively easy
to understand. Only interferences between $\pm1$ helicities and zero
helicities contribute to $\alpha_1$, while only interferences between
$\pm 1$ helicities and $\pm 1$ helicities contribute to $\alpha_2$. At
threshold, all helicity states are equally populated and the relative
interferences are maximized. For an off-shell $V$, the longitudinal mode
is slightly more populated than the transverse mode, reducing the
relative interference and thereby, the asymmetries. Above threshold, the
transverse modes are greatly suppressed relative to the longitudinal
mode, asymptotically like $(M_V/M_H)^2$, and so the asymmetries $\alpha_1$
and $\alpha_2$ fall like $(M_V/M_H)^2$ and $(M_V/M_H)^4$, respectively.

As may be seen from Eqns.(A17-20), the asymmetry coefficients are independent
of the final state  {\it except for} $\alpha_1(Z)$. $\alpha_1(Z)$ is maximized
when $Z^*Z^* \rightarrow $ two down-type quark pairs. For each up-type quark
pair in the final state $\alpha_1(Z)$ is reduced by 0.73; for each
charged lepton pair by a factor of 0.17. In fact it is easy to see that
$\alpha_1(V)$ is maximized when the $Vf \bar f$ coupling is purely chiral
(either left-handed or right-handed) so that parity violation is maximal,
and goes to zero for a coupling that is pure vector or axial vector.

The nearness of $\sin^2\theta_w$ to the ``magic" value 1/4 means that
the vector coupling of the $Z$ to a charged lepton pair is nearly zero; this
in turn means that the asymmetry $\alpha_1(Z)$ with a $l^+l^-$ in the
final state is nearly zero. This circumstance is most unkind, since the
$Z \rightarrow l^+ l^-$ mode has the best signature, both in terms of
beating backgrounds, and beating the $\pm \hat n$ ambiguity.
Prospects for reconstruction of the Higgs and its decay planes are better
at $e^+ e^-$ machines than at hadron colliders.

As explained earlier, a CP-odd pseudoscalar meson does not couple to
the longitudinal $V$'s. Accordingly, it has a vanishing $\alpha_1$, but
a maximal and negative $\alpha_2$ value of $-1\over 4$ [\dkr]. Such would
be the asymmetries for a technipion, for example.

Incidentally, it is just this angular orientation of decay planes that was
used to definitely determine the parity of the neutral pion thirty years
ago [\history]. Perhaps after extraordinary effort
physics history will repeat itself.  Or perhaps other observables, such as
the fermion--antifermion lab energy asymmetries $E_+/E_-$
or averaged rest frame polar angles [\duncan], will prove more useful.

\vskip 1in

\centerline{4. HIGGS DECAY TO FOUR FERMIONS INCLUDING SINGLE TOP}

The coupling of the standard model Higgs to a fermion scales as the fermion
mass over the electroweak VEV, $\sim 250 \; \GeV$.
Therefore, unless a fermion is heavy
on the scale of the $W$ or $Z$, its coupling to the Higgs is negligible.
Here we
consider the modification of our decay rate to four fermions when fermion
masses are included.  We take one fermion to be massive, and continue to
neglect the other fermion masses [\inverse].  The  relevant example
we consider is the decay of a Higgs with mass above single top threshold but
below $t \bar t$ threshold. The decay mode is
$H \rightarrow$ top (mass $m_t$) plus three light
(on the scale of the $W$ boson, the $b-$quark qualifies) fermions.
In models where the Higgs is generated dynamically as a scalar
$t \bar t$ bound state [\ttbar], or where the top and Higgs masses
are tuned to cancel divergences [\div],
the Higgs mass is expected to lie in this
range $[m_t, 2m_t]$. We also note that the
branching ratio of toponium to a Higgs in
this mass range is expected to be a few percent via the
$H\gamma$ channel.  However, toponium is not expected to exist
if $m_t$ exceeds about 140
Gev, since then the top quark lifetime is shorter than the bound state
formation time.  Thus, if the $t\bar t$ threshold exceeds the Higgs mass,
then $H\rightarrow$ single top may offer the best hope for extracting the
$H t \bar t$ coupling.

In addition to the graph of Fig.1(a) already
considered, there is the additional graph of Fig.1(b), proportional to the
H$t\bar t$ coupling. Since $Z$ bosons
conserve flavor and therefore cannot produce a single top, the only
gauge boson contributing in Fig.1(a) is the $W$.  For improved accuracy,
we keep the mass of
the $b$-quark in the phase space, although it is omitted in the matrix
element.

The expression for the width is presented
in Eqns.(A.1)-(A.6).  The width for the process $H\rightarrow WW\rightarrow
t\bar b s \bar c + \bar t b \bar s c$
for various values of $m_t/M_H$ is displayed in Fig.4.
Single but not double production of top requires
$0.5<m_t/M_H<1.0$, and various choices are indicated in the figure.
{}From direct CDF top mass limits [\cdf] and from radiative correction
limits [\mtop], $89\GeV <m_t<182\GeV$ at $95\% CL.$ In our analysis, we
relax the $CL$ and assume that $m_t<250\GeV$ absolutely.
For comparison, we include the massless limit ($m_t/M_H=0$)
and the $H\rightarrow b \bar b)$ width in Fig.4.
Were the Higgs mass above $t\bar t$ threshold, the $H\rightarrow t\bar t$
width would be the $b\bar b$ curve scaled up by $\beta^{3/2} (m_t/m_b)^2$,
where $\beta = \sqrt{1-4 m_t^2/M_H^2}$.
The massless limit describes, for example, the process
$H \rightarrow WW \rightarrow u \bar d s \bar c + \bar u  d \bar s  c$,
and is roughly 25\% of the $H\rightarrow WW$ width.
One can see that for $m_t/
M_H\simeq 0.5$ (precisely at $t\bar t$ threshold) the partial width for
$H\rightarrow t \bar b s \bar c + \bar t b \bar s c$
is reduced by only a factor of 10 or less compared to
the massless limit for $M_H \sim 200 - 500$ GeV.
For larger values of $m_t/M_H$, the partial width falls off rapidly.
The ratio $m_t/M_H$ determines how far off-shell the virtual top is, and so
determines the qualitative features of the rate.
Fig.5 amply demonstrates this strong sensitivity.  Starting
at $m_t/M_H=0.4$ ($M_H=400\GeV$), the rate for
$H\rightarrow t\bar b s \bar c
 + \bar t b \bar s c$ drops by ${\cal O}(10^3)$ at $m_t/M_H=0.55$.
For decreasing values of $M_H$ and fixed $m_t/M_H$, the drop is increasingly
more severe due to diminishing phase space.
If $m_t/M_H=0.6$
and $M_H>200\GeV$, $\Gamma(H\rightarrow t\bar b s \bar c+\bar t b\bar s c)/
\Gamma(H\rightarrow \tau^+\tau^-)={\cal O}(1)$, which implies a drastic
reduction, since $H\rightarrow \tau^+\tau^-$ is well known to
be a rare decay mode.

An important question to consider at this point is the relative magnitude
of the couplings that contribute to the partial width. For large $m_t$, one
would guess that the Yukawa contribution should be significant. This
relationship is indeed demonstrated in Fig.6; the three curves show
the separate partial width contributions arising from the
Yukawa ($Y$), gauge ($G$), and Yukawa + gauge ($Y+G$) terms including the
interference.
The two graphs of Fig. 1 are individually gauge invariant so the separation
into $Y$ and $G$ is meaningful.
In Fig.6, $m_t/M_H$ is chosen to be 0.6, so 89 GeV $\leq m_t \leq$ 250 GeV
confines $150\GeV \leq M_H \leq
420\GeV$ as the physically relevant region. One can observe that $\Gamma_{Y}$
is the dominant contribution for this range of $M_H$ by an
order of magnitude over $\Gamma_{G}.$
At fixed $M_H$, $\Gamma_{Y}$ will dominate even more for larger values of
$m_t/M_H$.

The structure of the curves can be understood as follows: In the gauge
amplitude (Fig.1a) the intermediate $W$ connecting to $(t\bar b)$ is
kinematically allowed to venture on-shell when $M_H>M_W>m_t+m_b$. In
Fig.6 this occurs for $M_W<M_H<125 \GeV$, and we
see structure in $\Gamma_{G}$ in this region. In the gauge and Yukawa
amplitudes the intermediate $W$ connecting the massless $q\bar q$ pair
is kinematically allowed on-shell when $M_H>M_W+m_t+m_b$. In Fig. 6 this
occurs for $M_H>215\GeV$, and we see a sharp rise in both amplitudes at
this value.

\vskip 1in

\centerline{5. SIGNATURE AND BACKGROUND}

The final states for $H\rightarrow t\bar b W^{-(*)}\rightarrow b W^{+(*)}
\bar b W^{-(*)}$ will be the same as those from the QCD process
$gg\rightarrow t\bar t\rightarrow b W^{+(*)}\bar b W^{-(*)}$, thus making
the $H\rightarrow t3f$ mode difficult to extract at a hadron collider. The
signal to background issue here is quite similar to the well known
difficulties in extracting an above-threshold $H\rightarrow t\bar t$
signal from a QCD background. In fact, within the top mass resolution
expected at LHC and SSC detectors [\paige], the $\bar b W^{-(*)}$ accompanying
the single top may well mimic an associated $\bar t$. With a $b W^{(*)}$/
associated-top misidentification, the signature difficulties are identical
for the $H\rightarrow$ single top and the $H\rightarrow t\bar t$ possibilities.
However, unlike the $H\rightarrow t\bar t$ branching ratio, the $H\rightarrow$
single top branching ratio is small, making signal extraction much more
difficult.

What may be fruitful at a hadron collider
is to trigger on $t\bar t$ produced in association
with Higgses. The dominant Higgs inclusive
production mechanism below $M_H=500\GeV$ is
gluon-gluon fusion, which yields a cross section of $\simeq 40$ pb at
SSC energies ($\sqrt{s}=40$ TeV), fairly flat for $M_H\leq 500 \GeV$. The
$Ht\bar t$ cross section at SSC energies, also dominated by the gluon-gluon
fusion mechanism, is $\simeq 4$ pb at $M_H=150 \GeV$, falling to $\simeq 0.3$
pb at $M_H=500 \GeV$; it is rather independent of the top mass in the
range of our interests. Thus the loss in event rate is a factor of only
$\simeq$ 10 to 130; on the order of $10^4$ $H t\bar t$
events per year may be expected
assuming the standard SSC integrated luminosity of $10^4$ pb$^{-1}$/SSC yr
($10^7$ sec). The $Ht\bar t$ cross section at the LHC ($\sqrt{s}=16$ TeV) is
down from that of the SSC by an order of magnitude or more (a function of
masses). However, the higher design luminosity at the LHC may argue for a
number of $Ht \bar t$ events comparable to the SSC. It has been
shown recently [\marcpaige] that triggering on
an isolated secondary lepton, $e$ or $\mu$,
from decay of the associated $t$ or $\bar t$, renders discovery of the
intermediate mass Higgs feasible at the SSC, via the rare decay mode
$H\rightarrow \gamma \gamma$, ($BR \simeq 10^{-3}$); in an analogous fashion,
one may hope that extraction of the rare decay mode presented here,
$H\rightarrow t \bar b W^{(*)}$, may also be feasible in association with a
$t\bar t\rightarrow$ isolated $l^\pm$ trigger; a $pp\rightarrow t\bar t
(H\rightarrow t \bar b W^{(*)})X$ event (or above $t\bar t$-threshold, the
$pp\rightarrow t\bar t(H\rightarrow t \bar t)X$ event) would lead to a
spectacular number of high $p_T$ jets and leptons.  If the Higgs is not
reconstructible in the single top mode, then an excess of spectacular
events may be the only signal from the mode.  Possibly helpful for signal
enhancement is the fact that W's from the Higgs decay chain will be
predominantly longitudinally polarized [\wlong], whereas ''background"
W's produced by light quarks will be transversely polarized.

The order $\alpha_s^4$ $pp\rightarrow t\bar t t\bar t$ QCD background to
our process has been calculated recently [\bspm]. The result is
$\sigma_{QCD}(pp\rightarrow t\bar t t\bar t X)=1, $
0.4, 0.2, and 0.07 pb at the
SSC for $m_t$ = 110, 140, 170, 200 $\GeV$; and 0.1, 0.03, 0.008, and 0.003 pb
for the same $m_t$ values, at the LHC. Thus, at either the SSC or the LHC,
$\sigma(pp\rightarrow t\bar t H X)/\sigma_{QCD}(pp\rightarrow t\bar t t\bar tX)
\simeq 0.3\; {\rm to}\; 50$, depending on the Higgs and top masses, in the
range of our interest ($100\;\GeV\leq M_H\leq 500\;\GeV, 90\;\GeV
\leq m_t\leq 250 \GeV$). Thus a $BR\geq 10^{-2}$ in the $H\rightarrow
tbW^{(*)}$ mode may in principle be observable.
{}From figure 4, if $M_H>160 \GeV$, then $m_t<0.55M_H$ appears to
be needed to obtain such a branching ratio. A concern is that there is no
obvious signature distinguishing the $t``\bar t"$ from $H\rightarrow t
bW^{(*)}$, from the second $t\bar t$ in the QCD signal; all tops will tend to
be produced near threshold. The extra combinatoric possibilities available
with four tops is also problematic. A Monte Carlo simulation is required
to quantitatively determine detection capabilities for the $H\rightarrow t
bW^{(*)}$ mode.

At a future $e^+e^-$ collider, it may prove possible to probe a smaller
$BR(H\rightarrow tbW)$, and hence a larger $m_t/M_H$ ratio, if the
integrated luminosity is sufficiently high, and if the $t\bar t$ and
$t\bar t t\bar t$ backgrounds are more controllable.
The associated Z+H and/or WW fusion cross section for production of
a 100 GeV SM Higgs is 0.30, 0.14, and 0.24 pb for the values
$\sqrt{s} = 200, 500$, and $1\;TeV$, while for a 200 GeV
Higgs it is roughly 0.1 pb over the range $\sqrt{s} = 500\;GeV$ to $\sqrt{s} =
1\;TeV,$ and for a 500 GeV Higgs it is 0.03 pb at $\sqrt{s} = 1\;TeV$
[\hunter].
Thus, with $10 fb^{-1}$/yr of luminosity, 300 to 3000 Higgs events per year are
anticipated.  A $100 fb^{-1}$/yr machine has even been discussed, for which
the rates are larger by a factor of ten.
We conclude that branching ratios as small as $10^{-3}$ may be detectible.

Finally, we comment that if $M_H$ exceeds $t\bar t$ threshold, as opposed
to the mass range $m_t<M_H<2m_t$ considered in this paper, then
$$
{{\sigma(pp\rightarrow t\bar t(H\rightarrow t\bar t)X)}\over
{\sigma_{QCD}(pp\rightarrow t\bar t t\bar tX)}}
\sim (0.3 \; {\rm to} \; 50)\times BR(H\rightarrow t\bar t), \eqno(4)
$$
which suggests that the $Ht\bar t$ coupling may be experimentally accessible if
$BR(H\rightarrow t\bar t)$ exceeds $10^{-2}$. Using SM formulae, one finds
$BR(H\rightarrow t\bar t)\simeq \Gamma(H\rightarrow t\bar t)/\Gamma(H
\rightarrow t\bar t,ZZ,W^+W^-)=r(1-r)^{3\over 2}/[1+r(1-r)^{3\over 2}]$,
where $r\equiv 4m_t^2/M_H^2$, and terms of order $M_Z/M_H$ have been
neglected. This branching ratio can be as large as $30\%$ (at $r=0.4$, or
$m_t\sim M_H/3$), clearly warranting further study of this mode. As with
the sub-threshold process $H\rightarrow t X$, a Monte Carlo simulation is
required to determine signal viability.

\vskip 1in

\centerline{6. FERMIOPHOBIA, GAUGEOPHOBIA, AND FERMIOPHILIA}

Given the possibility that gauge boson masses and fermion masses may arise
from different mechanisms, one can imagine scalar fields (i) with couplings
to fermions suppressed relative to the SM values $m_f/v_{EW}$, (ii) with
couplings to $WW$ and $ZZ$ suppressed relative to the SM values
$g^2v_{EW}/ 2$, and $g^2v_{EW}/(2\cos^2\theta_w)$,
respectively, and
(iii) with enhanced couplings to the fermions. We will label these scalars
fermiophobic, gaugeophobic, and fermiophilic. One can easily invent
models of each kind, and find examples of each kind in the literature. We have
already mentioned  the $\phi \rightarrow - \phi$ fermiophobic scalar.
Any scalar in an $SU(2)$ representation other than doublet will also be
fermiophobic, since the $SU(2)$ invariance then forbids the
$\phi \bar \Psi _L \Psi_R$ operator.

The $\phi^2 WW, \phi^2 ZZ, (\phi^{\ast} \overlrarrow{\partial}
\phi)W,$ and $(\phi^{\ast} \overlrarrow{\partial}
\phi)Z$ couplings are fixed by the gauge coupling
$g$. However, the $\phi WW$ and $\phi ZZ$ couplings are proportional to
$g^2<\phi>$. Accordingly, the latter are suppressed if
$\sqrt 2 <\phi>$ is less than $v_{EW} = 246$ GeV.
In fact, the measured value of the
$\rho$-parameter tells us that VEVs of representations  larger than doublet
are much smaller than $v_{EW}$, so scalars from
representations larger than doublet
are gaugeophobic as well as fermiophobic. Higgs doublets may themselves be
gaugeophobic, since in a multi--doublet model, $v_{EW}^2 = \sum_{\rm doublets}
v_D^2$ sums the positive contribution from each doublet. Further models
abound. As mentioned earlier, in models which conserve CP in the Higgs and
gauge--boson sectors, the pseudoscalar Higgses are gaugeophobic [\hunter]. In
general two--doublet models, the charged Higgs and the single CP--odd Higgs
are gaugeophobic [\hunter].

Fermiophilic scalars may result from multi--scalar mixing. A well known
example is the general two--doublet model, in which some Yukawa
couplings are enhanced by VEV ratios, while others are suppressed
(fermiophobia again) by the inverse ratio. (Particularly well motivated
two--doublet models are the minimal supersymmetric model invented to
stabilize the weak scale, and the Peccei--Quinn model [\pq],
invented to solve the strong CP problem.) There is also the possibility
of scalar fields in representations which do not get a VEV. These
non--Higgs scalars would have no tree--level $\phi WW$ or $\phi ZZ$
couplings (i.e. gaugeophobic) and arbitrary Yukawa couplings (fermiophobic
or fermiophilic). Such ``extra" scalars have been invoked in many
theoretical contexts [\nonhiggs].

Finally, we mention hybrid models where the VEV of a fundamental scalar
field generates one mass scale, while other masses arise from a
condensate or from mixing with heavy singlet fermions [\simmons].
Such models incorporate the
best attributes of technicolor or fermion singlet masses
with the simple attributes of Higgs physics,
but with an increase in field content and a decrease in aesthetics.
What is clear is that when a Higgs particle is finally discovered, its
branching ratios will be very revealing. From the SM branching ratios of
Fig.2 in reference [\BARGER], one may infer the dominant decay modes of
non--standard Higgses: for a neutral fermiophobic Higgs, omit the
two--fermion and two--gluon modes; for a neutral gaugeophobic Higgs, omit the
four--fermion, $WW$, and $ZZ$ modes. Ignoring the loop-and mixing-induced
$H \rightarrow f \bar f$ rate and the two--loop $H \rightarrow gg$ rate,
one learns that the dominant decay mode for a fermiophobic Higgs is $\gamma
\gamma$ for $M_H \leq 80$ GeV, and four fermions for $M_H>80\; \GeV$
(via $W^*W^*$ for 80 GeV $\leq M_H \leq$ 100 GeV, via $W^*W$ for
100 GeV $\leq M_H \leq 2M_W$, and via $WW$ above the production threshold).
Estimates of the rates for the induced $\bar f f$ and $gg$ modes require
calculations within a specific model.  For a heavy gaugeophobic Higgs, the
$t \bar{t}$ mode is dominant over the WW mode.

At the SSC machine with energy $\sqrt s \sim 40$ TeV,
the $gg \; \rightarrow$ top-loop
$\rightarrow H$ chain is expected to be the dominant production
mechanism for a standard Higgs with $M_H \leq 5m_t$ (as here). However, a
fermiophobic Higgs may in fact have as its dominant production mode
$WW$ fusion, or $W^*\rightarrow W H$, with a smaller production rate.
At an $e^+ e^-$ machine, the standard Higgs is expected to be produced via
$Z^* \rightarrow Z+H$ for $\sqrt{s}$ up to $400 GeV + 0.6 M_H,$ and by WW
fusion at higher $\sqrt{s}$ [\hunter].  A gaugeophobic Higgs would have a
suppressed production rate, either in association with $b {\bar
b}\;\rm{or}\;t \bar{t}$ (if kinematicaly allowed), or via a top-loop.

Let us now discuss changes in the rate for $H \rightarrow t + $ 3 fermions
when the Higgs is non--standard. Let $v/ \sqrt 2$ denote the VEV of the
non--standard scalar multiplet. Then the $HWW$ coupling is $g^2v/2$
and the $H \bar t t$ coupling is an arbitrary constant $g_Y$. If we
make the reasonable assumption that all Higgs fields couple to top with the
same sign, then $m_t = \sum_{\rm scalars} g_Y v / \sqrt 2$ provides the
constraint $g_Y < \sqrt 2 m_t/v$; we write $g_Y = \sqrt2 \xi m_t / v$,
$0 \leq \xi \leq 1$. If $v < v_{EW},$ the Higgs is gaugeophobic; if
$\xi / v < 1 / v_{EW}$, the Higgs is fermiophobic; and if $\xi / v > 1 /
v_{EW}$, the Higgs is fermiophilic. If $H$ is in fact a scalar without a VEV,
then setting $v$ equal to zero yields the appropriate result: decoupling
from $W^+W^-$ and an arbitrary coupling to $\bar t t$.
The rate for non--standard $H \rightarrow t +$ 3 fermions in terms of
$v$ and $\xi$ can be obtained from Fig.6:
$\Gamma_{Y}$ scales as $(\xi v_{EW}/v)^2$,
$\Gamma_{G}$ scales as $(v/v_{EW})^2$, and
the interference term scales simply as $\xi$.
For a truly gaugeophobic Higgs, $\Gamma_{Y}$ $\underline{is}$ the total
width, while for a truly fermiophobic Higgs, $\Gamma_{G}$ $\underline{is}$
the total width.

\vskip 1in

\centerline{7. SUMMARY}

We have presented the tree-decay of the SM Higgs to four fermions,
including the possibility of a single massive fermion in the
final state in order to explore
the potentially large Yukawa contribution to the decay rate.
In the massless limit, we have
demonstrated that if both gauge bosons are off-shell, the rate is considerably
reduced. If one gauge boson can go on--shell, it will, and
therefore  the OGBOS approximation is valid for an intermediate
mass Higgs. We have also
explored how a measurement of the angular correlation of the decay planes of
the final state fermions may lead to a determination of the intrinsic parity
of the Higgs, even if $M_H<2M_W$.

Next, we have shown that for a Higgs mass below $t \bar t$
threshold, the decay rate to single top is generally
dominated by the Yukawa coupling, and thus tests the SM mechanism for fermion
mass generation. For a Higgs just below the
$t\bar t$ threshold, the single top rate is only
${\cal O}(10^{-1})$ down from the
massless mode where the gauge coupling is the only contribution. As $m_t/M_H$
grows, the overall rate is dramatically reduced.
It is evident (see e.g. Fig.4) that the
$H\rightarrow t\bar b W^{(*)}$ process remains
competitive in rate with other potentially detectable
rare decay modes such as $H\rightarrow \gamma \gamma,
Z\gamma,b\bar b,\tau^+\tau^-$ over much of the allowed range of $M_H$,
$m_t/M_H$.
At a hadron collider, the signal to background is enhanced by triggering on
$pp\rightarrow t\bar tHX$ signatures, making $H\rightarrow$ single top
possibly observable if $2m_t {<\atop \sim} 1.1\;M_H$;
a Monte Carlo simulation is needed to quantitatively compare the signal to the
${\cal O}(\alpha^4_s)\; t\bar t t\bar t$ QCD background. (Prospects appear
brighter for measuring $\Gamma(H\rightarrow t\bar t)$ via $pp\rightarrow
t\bar tHX$, should $M_H$ exceed $2m_t$.)  It appears that $H \rightarrow$
single top may have a better chance for detection at a future high energy,
high luminosity $e^+ e^-$
collider, where branching ratios as small as $10^{-3}$, i.e.
$2m_t {<\atop \sim} 1.2\;M_H,$ may be measureable.

Finally, we have shown that the rate for Higgs $\rightarrow$ single top
is a sensitive measure of any ``non-standardness'' in either the
Higgs-gauge boson and/or the Higgs-fermion sectors of the true theory.
We havediscussed the implications of ``non-standardness''
in these sectors for the
general issue of fermion and gauge boson mass generation.

\endpage

\ack


We have benefitted from discussions with Bill Marciano, Jack Gunion,
Howie Haber and Wendell Holladay.
This work was supported in part by U.S. Department of
Energy Grants No. DE-FG05-85ER40226, DE-AC02-76-ER022789, and
DE-FG03-84ER40168.
Part of this work was carried out at the Aspen Center for Physics.

\endpage

\centerline{APPENDIX}

The Lorentz invariant matrix element for particle decay to a four body final
state depends on five independent variables.  We find it convenient to use as
variables the invariant mass of each fermion pair connected by a common fermion
line (there are two of these), the polar angle of the particle momentum with
respect to the pair momentum direction, evaluated in the pair center of mass
frame (there are two of these), and the azimuthal angle describing the
orientation of one plane defined by paired fermion momenta relative to the
plane defined by the other paired momenta (there is one of these).  For the
case with all massless final state particles, where only graph 1(a)
contributes, the first four of these variables are just the virtual $W$ or
$Z$
boson invariant masses, and the fermion momentum direction relative to its
parent $W$ or $Z$ momentum direction, evaluated in the $W$ or $Z$
boson rest frame.
The Lorentz invariant phase space in terms of these variables is
$$
\eqalignno{\int dlips^{(4)} =  \int_{(m_t+m_b)^2}^{M_H^2} dQ^2_1 &
     \int_{0}^{(M_H-{\sqrt {Q_1^2}})^2} dQ^2_2
     \int_0^{2\pi} d\phi
     \int_{-1}^{1} d\cos\theta_1
     \int_{-1}^{1} d\cos\theta_2 \cr
   & \cdot {\lambda^{1/2}(M^2_H,Q^2_1,Q^2_2)\lambda^{1/2}(Q^2_1,m^2_t,m^2_b)
     \over 2^8M_H^2Q^2_1(2\pi)^6} \quad , &(A.1)\cr}
$$
where $\lambda (a,b,c)=a^2+b^2+c^2-2(ab+bc+ca)$ is the usual triangle
function.
(An alternate choice of the five independent phase space variables is
discussed in [\KWMT]).  The width is
$$
\Gamma(H\rightarrow (t{\bar b} \;+ \; {\bar t} b)\,+\,f{\bar f})
={1 \over M_H}\int dlips^{(4)} \bigg[ \sum_{spin} \vert {\cal M} \vert^2
\bigg] \quad , \eqno(A.2)
$$
where
$$
\sum_{spin} \vert {\cal M} \vert^2 = \sum_{spin}
 \bigg[ {\cal M}_a{\cal M}_a^{\ast}+{\cal M}_b{\cal M}_b^{\ast}+
  2Re{\cal M}_a{\cal M}_b^{\ast} \bigg] \quad , \eqno(A.3)
$$
expresses the summation over the squares of the amplitudes from
Figs.1(a) and 1(b), and the
interference of the two amplitudes.  The expressions for these
squared matrix elements are
$$
\eqalignno{\sum_{spin} {\cal M}_a{\cal M}_a^{\ast} & =
    3^ng^6 M_W^2 B_W(Q^2_1) B_W(Q^2_2) \cr
    &  \cdot \left\{ 4f\cdot t\, {\bar f}\cdot \bar b\,-2({m_t \over M_W})^2
      \left( {\bar f}\cdot \bar b\,f\cdot Q_1+
       f\cdot \bar b \,{\bar f}\cdot Q_1 -
       {1 \over 4}Q_2^2(Q_1^2-m_t^2) \right)  \right. \cr
    & + \left. {m_t^2(Q_1^2-m_t^2) \over M_W^4}
            \left( f\cdot Q_1\,{\bar f}\cdot Q_1
      - {1 \over 4}Q_1^2Q_2^2 \right) \right\} \quad , &(A.4)\cr}
$$
$$
\sum_{spin} {\cal M}_b{\cal M}_b^{\ast}  =
  {3^ng^6m_t^2 \over M_W^2} B_W(Q^2_2)
\bigg[(M_H^2-2P\cdot t)^2+(m_t\Gamma_t)^2\bigg]^{-1}
     \cdot \left\{ {\bar f}\cdot \bar b\,
            \left( 2f\cdot P\,t\cdot P\,-M_H^2\,f\cdot t \right)
      \right\} \quad , \eqno(A.5)
$$
$$
\eqalignno{2 Re\sum_{spin} {\cal M}_a{\cal M}_b^{\ast} & =
  -2 {3^ng^6 m_t^2 \over M_W^2}
\bigg[(M_H^2-2P\cdot t)^2+(m_t\Gamma_t)^2\bigg]^{-1} B_W(Q^2_1) B_W(Q^2_2) \cr
  &  \cdot \biggl\{ \bigg[(M_W^2-Q_1^2)(M_H^2-2P\cdot t)+
M_W\Gamma_Wm_t\Gamma_t\bigg]
         \bigg[ 2M_W^2f\cdot P\,{\bar f}\cdot \bar b- \cr
  &       Q_1\cdot {\bar f} \left( P\cdot Q_1\,f\cdot \bar b\,
-P\cdot \bar b\,Q_1\cdot f
                    \right) \cr
  &      -Q_1\cdot \bar b \left( P\cdot {\bar f}\,Q_1\cdot f\,
-P\cdot Q_1\,f\cdot
           {\bar f} \right) \cr
   &  -{Q_1^2 \over 2} \left( P\cdot f\,{\bar b}\cdot {\bar f}\,
    +P\cdot \bar b\,f\cdot {\bar f}\,-P\cdot {\bar f}
\,\bar b\cdot f \right) \bigg] \cr
  &  +{1 \over 2}\bigg[ M_W\Gamma_W(M_H^2-2P\cdot t)+
m_t\Gamma_t(Q_1^2-M_W^2)\bigg] \left( 2Q_1\cdot {\bar f}\, + Q_1^2 \right)\cr
  &    \epsilon^{\alpha\beta\mu\nu}P_{\alpha}{\bar f}_{\beta}Q_{1\mu}
      {\bar b}_{\nu} \biggr\}  \quad , }
$$
\rl{(A.6)}

\noindent
where $g = \vert e \vert/\sin\theta_w$,
and $B_V(Q^2) = [(Q^2-M_V^2)^2+M_V^2\Gamma_V^2]^{-1}$ is the vector boson
Breit-Wigner factor ($V$ = $W$ or $Z$).
For convenience, we take $M_Z$ and $\Gamma_Z$ to be $91$ and $2.5$ GeV,
respectively, and $M_W$ and $\Gamma_W$ to be $80$ and $2.1$ GeV,
respectively. The weak mixing angle is taken to be $\sin^2\theta_w = 0.23$.
In all of our rate equations, $3^n$ is a color factor, with $n$ being the
number of quark-antiquark pairs in the final state.

Note that
we have retained the $b$-quark mass in the phase space.  However, we have
omitted it in the above matrix elements since to the same order in
$m_b/M_W$ there
is a further graph given by exchanging the $b$ and $t$ quarks in Fig.1(b),
which we have not included. We take $m_b = 4.7 \GeV$.
For the top width $\Gamma_t$, we have summed the
following expression over $(f,\bar f')=(u,\bar d),(c,\bar s),(\nu_e,e^+),
(\nu_\mu,\mu^+),(\nu_\tau,\tau^+):$

$$
\Gamma(t\rightarrow bW^{(*)}\rightarrow b f \bar f')
={3^nm_tg^4 \over 3\pi^3\cdot 2^{10}}\int_{2\sqrt{r}}^{1+r}
dx_b {{\sqrt{{x_b}^2-4r}(3(1+r)x_b-2{x_b}^2-4r)}\over
((1-x_b+r)-{\hat M}_W^2)^2+
{\hat \Gamma}_W^2{\hat M}_W^2} \eqno(A.7)
$$

\noindent
where $x_b=2E_b/m_t, r=m_b^2/m_t^2$, and ${\hat M}_W,{\hat \Gamma}_W=M_W/m_t,
\Gamma_W/m_t$ respectively.

It is not difficult to analytically integrate out the light fermion pair.
The resulting expression depends on three variables, which we choose to be the
paired fermion invariant masses (two of these) and twice the dot product of
the Higgs four-momentum ($P$) with the heavy fermion four-momentum ($t$);
we call this invariant variable
$\xi =2P\cdot t$ .  The Lorentz invariant phase space is
$$
\int dlips^{(4)}={1 \over 2^6(2\pi)^5M_H^2} \int_{(m_t+m_b)^2}^{M_H^2}
      dQ_1^2 \int_0^{(M_H-{\sqrt {Q_1^2}})^2} dQ_2^2
      \int_{\xi_-}^{\xi_+} d\xi \quad , \eqno(A.8)
$$
with
$$
\xi_{\pm}={1 \over 2Q_1^2} \bigg[ (M_H^2+Q_1^2-Q_2^2)(Q_1^2+m_t^2-m_b^2)
                 \pm \lambda^{1/2}(Q_1^2,m_t^2,m_b^2)
                      \lambda^{1/2}(M_H^2,Q_1^2,Q_2^2) \bigg] \; \; .
$$
\rl{$(A.9)$}
A technical difficulty arises as this point. As $m_b$ is turned on from zero,
$\xi_-$ decreases, therein going out of the physical region for the
$m_b^2=0$ matrix element. We have cured this disease by setting
$\xi_-\equiv \xi_-(m_b=0),\xi_+\equiv \xi_-(m_b=0)+
[\xi_+(m_b\not= 0) - \xi_-(m_b\not= 0)]$.

We find for the squared matrix element, in the $m_b=0$ limit,
$$
\eqalignno{\sum_{spin} {\cal M}_a{\cal M}_a^{\ast} & =
    {3^n\pi g^6M_W^2 \over 24} B_W(Q^2_1) B_W(Q^2_2) \cr
    &  \cdot \biggl\{ 2Q_2^2(2Q_1^2-\xi)-
          2(\xi-Q_1^2-m_t^2)(\xi-M_H^2-m_t^2) \biggr. \cr
    & + {2m_t^2 \over M_W^2} \bigg[ (M_H^2+m_t^2-Q_2^2-\xi)
           (Q_1^2+Q_2^2-M_H^2)
      +2Q_2^2(Q_1^2-m_t^2) \bigg] \cr
    & +\left. {m_t^2 \over 2M_W^4} (Q_1^2-m_t^2) \lambda (M_H^2,Q_1^2,Q_2^2)
           \right\} \; \; , &(A.10) \cr}
$$
$$
\eqalignno{\sum_{spin} {\cal M}_b{\cal M}_b^{\ast} & =
  {3^n\pi g^6m_t^2 \over 48 \,M_W^2} B_W(Q^2_2)
\bigg[ (M_H^2-\xi)^2+(m_t\Gamma_t)^2\bigg]^{-1} \cr
 &  \cdot \biggl\{ \xi \bigg[
   (Q_2^2-Q_1^2)(2M_H^2+m_t^2-2Q_2^2)-M_H^2m_t^2 \bigg] \cr
  &    + M_H^2 \bigg[ (M_H^2+m_t^2)(Q_1^2+m_t^2)-2Q_2^2Q_1^2 \bigg]
   + \xi^2(Q_1^2-2Q_2^2) \biggr\}  \quad , }
$$
\rl{(A.11)}

$$
\eqalignno{2Re\sum_{spin} {\cal M}_a{\cal M}_b^{\ast} & =
   -{\pi \over 12}3^ng^6m_t^2
\bigg[ (M_H^2-\xi)^2+(m_t\Gamma_t)^2\bigg]^{-1}
\bigg[(M_W^2-Q_1^2)(M_H^2-\xi)+M_W\Gamma_Wm_t\Gamma_t\bigg]\cr
  & B_W(Q^2_1) B_W(Q^2_2) \cdot \biggl\{ \xi \left( Q_1^2-2Q_2^2-M_H^2+{Q_1^2
                 \over 2M_W^2}(M_H^2+Q_2^2-Q_1^2) \right) \cr
  & + {1 \over 2M_W^2} \bigg[ -M_H^2Q_1^2(M_H^2-Q_1^2-Q_2^2)
      +m_t^2 \left( (Q_1^2-Q_2^2)^2-M_H^2(Q_1^2+Q_2^2) \right) \bigg] \cr
  & + M_H^2(M_H^2+m_t^2)+(Q_1^2-Q_2^2)(2Q_2^2-M_H^2-m_t^2)\biggr\}
                                \quad . }
$$
\rl{(A.12)}

\noindent
The limit of all massless final state particles is easily obtained from
the above equations.
For the $V^{\ast}V^{\ast}$ intermediate state $(V=W \; {\rm or} \; Z)$,
the result is
$$
\eqalignno{\Gamma (H \rightarrow & V^* V^* \rightarrow
f_1{\bar f}^{\prime}_1f_2{\bar f}^{\prime}_2)   \cr
 =  {4\cdot 3^ng^6M_V^2 \over 2M_H}
\int dlips^{(4)}
\biggl\{ C_L \left( f_1 \cdot f_2 \right) \;
             \left( {\bar f}^{\prime}_1 \cdot {\bar f}^{\prime}_2 \right)
 + & C_R \left( f_1 \cdot {\bar f}^{\prime}_2 \right)
        \; \left( f_2 \cdot {\bar f}^{\prime}_1 \right)
             \biggr\} \cr
  & \cdot B_V(2f_1\cdot \bar {f_1}') B_V(2f_2\cdot \bar {f_2}')}
$$
\rl{(A.13)}

\noindent
For the $V=W$ mode, the chiral couplings are just
$$
C_L=1 \; \; , \qquad \qquad \; C_R=0 \; \; . \eqno(A.14)
$$
For the $V=Z$ mode, they are
$$
\cos^6\theta_w C_L = {1 \over 2}(v^2_1+a^2_1)(v^2_2+a^2_2)+
2v_1v_2a_1a_2 \; \; , \eqno(A.15)
$$
and
$$
\cos^6\theta_w C_R = {1 \over 2}(v^2_1+a^2_1)(v^2_2+a^2_2)-
2v_1v_2a_1a_2 \; \; , \eqno(A.16)
$$
with $v_i=(T_{3R}+T_{3L}-2Q\sin^2\theta_w)_i, \; a_i=(T_{3R}-T_{3L})_i,$
where $T_{3L}$ and $T_{3R}$ are the weak isospin eigenvalues of the
left and right helicity fermions, and $Q$ is the fermion electric
charge in unit of $\vert e \vert$.
Integrating these expressions over the two polar angles leads to the following
expressions for the asymmetry parameters (defined in Eqn.(1) in the main
text):
$$
\eqalignno{\alpha_1(V)=\left( {C_R-C_L \over C_R+C_L} \right)
 {9\pi^2 \over 32\, D}
  \int_0^{M_H^2} dQ_1^2 & \int_0^{(M_H-{\sqrt {Q_1^2}})^2} dQ_2^2
  \lambda^{1/2}( M_H^2,Q_1^2,Q_2^2) B_V(Q^2_1) B_V(Q^2_2) \cr
  &  \cdot (M_H^2-Q_1^2-Q_2^2){\sqrt {Q_1^2Q_2^2}} \; \; \; , &(A.17)\cr}
$$
$$
\alpha_2(V)={1 \over  D}
  \int_0^{M_H^2} dQ_1^2 \int_0^{(M_H-{\sqrt {Q_1^2}})^2} dQ_2^2
  \lambda^{1/2}(  M_H^2,Q_1^2,Q_2^2)
   B_V(Q^2_1) B_V(Q^2_2) Q_1^2 Q_2^2 \; \; \; ,\eqno(A.18)
$$
with
$$
D={1 \over  2}
  \int_0^{M_H^2} dQ_1^2 \int_0^{(M_H-{\sqrt {Q_1^2}})^2}
 dQ^2_2  \lambda^{1/2}(  M_H^2,Q_1^2,Q_2^2)
B_V(Q^2_1) B_V(Q^2_2)
   \bigg[ 8Q_1^2Q_2^2+(M_H^2-Q_1^2-Q_2^2)^2 \bigg] \quad . \eqno(A.19)
$$
The final state dependence of the asymmetry $\alpha_1(Z)$ is given by
$$
\left( {C_R-C_L \over C_R+C_L} \right) = - {4v_1v_2a_1a_2 \over
(v_1^2+a_1^2)(v_2^2+a_2^2)} \quad . \eqno(A.20)
$$

In the massless fermion limit, further integration of the general expression
(A.13) may be done analytically,
leading to
$$
\eqalign{\Gamma  (H\rightarrow & V^{\ast}V^{\ast}\rightarrow f_1{\bar f}_1'
f_2{\bar f}_2') = {g^6m^2_V3^n \over 2^9\cdot 9\cdot M_H^3 (2\pi)^5}
[C_L+C_R] \cr
& \int^{M_H^2}_0 dQ_1^2 \int_0^{(M_H-{\sqrt {Q_1^2}})^2} dQ_2^2
\lambda^{1/2}(M_H^2,Q_1^2,Q_2^2)[8Q_1^2Q_2^2+(M_H^2-Q_1^2-Q_2^2)^2]
 B_V(Q_1^2) B_V(Q_2^2) \quad . \cr}
$$
\rl{$(A.21)$}

When kinematics allow one or both vector boson(s) to go on-shell, it will.
Then the appropriate limits of our general expressions are found by
applying the NWA, $B_V(Q^2)\rightarrow \pi\delta(Q^2-M_V^2)/M_V\Gamma_V$
(times a possible symmetry factor of 2, as explained in the text). As a check
on formulae (A.17) to (A.21), we apply the NWA to each vector boson and
obtain known on-shell results. The on-shell values of the asymmetries of
$\alpha_1(V)$ and $\alpha_2(V)$ are given in the text; the on-shell values of
Eqn.(A.21) are just
$$
\Gamma(H\rightarrow W^+W^-\rightarrow f_1\bar f'_1 f_2\bar f'_2)=
\Gamma(H\rightarrow W^+W^-) BR(W^+\rightarrow f_1\bar f'_1)
BR(W^-\rightarrow f_2\bar f'_2) \quad .
\eqno(A.22)
$$

\noindent
and
$$
\Gamma(H\rightarrow ZZ\rightarrow f_1\bar f_1 f_2\bar f_2)=
2\Gamma(H\rightarrow ZZ) BR(Z\rightarrow f_1\bar f_1)
BR(Z\rightarrow f_2\bar f_2) \quad .
\eqno(A.23)
$$

Similarly, the NWA replacement $B_V(Q^2)\rightarrow
\pi\delta(Q^2-M_V^2)/M_V\Gamma_V$ in Eqns.(A.1-6) and
(A.8-12) yields the rate for
$\Gamma(H \rightarrow t {\bar b}W) BR(W \rightarrow f{\bar f}')$.

\endpage

\centerline{FIGURE CAPTIONS}

1. Feynman diagrams for Higgs decay to four fermions. (a) is the gauge coupling
contribution, (b) is the Yukawa contribution in the case of a massive fermion.
$V$ is a generic symbol for either $W$ or $Z$ gauge bosons.

2. Differential decay rate for $H\rightarrow u{\bar d}s{\bar c} +
\bar u d \bar s c$
versus the scaled $W$ invariant mass $M_{inv}/M_W$, for various scaled Higgs
masses $M_H/M_W$.

3. Asymmetry parameters $\alpha_i(V)$, $(i=1,2),V=Z,W$, for the decays
$H \rightarrow Z^*Z^* \rightarrow d \bar d s \bar s$, and
$H \rightarrow W^* W^* \rightarrow u{\bar d}s{\bar c}$
versus the Higgs mass $M_H$. Only $\alpha_1(Z)$ is dependent on the final
state; for $u \bar u c \bar c \; (e^+e^-\mu^+\mu^-)$ it is lower than
$d \bar d s \bar s$ by factor of 0.53 (0.029).

4. $\Gamma(H\rightarrow t \bar b s \bar c + \bar t b  \bar s c)$
and the massless limit
$\Gamma(H\rightarrow u \bar d s \bar c + \bar u d \bar s c)$
versus $M_H$ for constant values of $m_t/M_H$.
The latter is roughly 25\% of the $H\rightarrow WW$ width.
The dotted lines correspond to regions excluded
by the standard model top quark bounds $89\GeV < m_t < 250\GeV$.
The dashed line corresponds to $\Gamma(H\rightarrow b\bar b)$; we have
not incorporated QCD corrections here, so the true $b\bar b$
width may be reduced as much as 50\% [\hunter].

5.  $\Gamma(H\rightarrow t \bar b s \bar c + \bar t b  \bar s c)$
versus $m_t/M_H$, for fixed $M_H=200,$ 300, 400 $\GeV$. The sharp
drop in $\Gamma_H$ across the $H\rightarrow t \bar t$
threshold ($m_t/M_H=0.5)$ for
increasing values of $m_t/M_H$ is due to increasing top virtuality.
In the range $2 m_t < M_H$ the width shown is just equal to $2\Gamma
(H\rightarrow t \bar t )/BR(t\rightarrow b \bar s c)$.

6. $\Gamma_i(H\rightarrow t \bar b s \bar c + \bar t b  \bar s c)$
 versus $M_H$, where $i=Y,G,$ and $Y+G$ refer to the Yukawa,
gauge, and total (including the interference term) contributions
to the partial width respectively;
here, $m_t/M_H=0.6$.

\endpage

\par\penalty-400\vskip\chapterskip\spacecheck\referenceminspace
   \ifreferenceopen \Closeout\referencewrite \referenceopenfalse \fi
   \line{\fourteenrm\hfil REFERENCES\hfil}\vskip\headskip
   \input referenc.txa
   
\bye